\documentclass[aps,prx, showkeys,reprint,superscriptaddress,floatfix,hidelinks, nofootinbib]{revtex4-2}

\usepackage[T1]{fontenc}
\usepackage[utf8]{inputenc}
\usepackage{graphicx}
\usepackage{hyperref}
\usepackage{xcolor}
\usepackage{siunitx}
\usepackage[version=4]{mhchem}
\usepackage{zref}
\usepackage{zref-xr}
\usepackage{zref-user}
\makeatletter
\newcommand*{\addFileDependency}[1]{
  \typeout{(#1)}
  \@addtofilelist{#1}
  \IfFileExists{#1}{}{\typeout{No file #1.}}
}
\makeatother
\newcommand*{\myexternaldocument}[1]{%
    \zexternaldocument{#1}%
    \addFileDependency{#1.tex}%
    \addFileDependency{#1.aux}%
}
\myexternaldocument{supplement-static}

\usepackage[nogroupskip,nonumberlist]{glossaries}
\setacronymstyle{long-short}
\newacronym{se}{SE}{Stokes-Einstein}
\newacronym{ses}{SES}{Stokes-Einstein-Sutherland}
\newacronym{svs}{SvS}{Svedberg-Stokes}
\newacronym{gk}{GK}{Green-Kubo}
\newacronym{pmf}{PMF}{potential of mean force}
\newacronym{com}{COM}{center of mass}
\newacronym{msd}{MSD}{mean square displacement}
\newacronym{rdf}{RDF}{radial distribution function}
\newacronym{acf}{ACF}{autocorrelation function}
\newacronym{vacf}{VACF}{velocity autocorrelation function}
\newacronym{facf}{FACF}{force autocorrelation function}
\newacronym{fsacf}{F\textsubscript{S}ACF}{stochastic force autocorrelation function}
\newacronym{ftacf}{F\textsubscript{T}ACF}{total force autocorrelation function}
\newacronym{vdw}{vdW}{van der Waals}
\newacronym{fdt}{FDT}{fluctuation-dissipation theorem}
\newacronym{psv}{PSV}{partial specific volume}
\newacronym{md}{MD}{molecular dynamics}
\newacronym{lj}{LJ}{Lennard-Jones}
\newacronym{wca}{WCA}{Weeks-Chandler-Andersen}
\newacronym{pme}{PME}{Particle mesh ewald}
\newacronym{NH}{NH}{Nos\'{e}-Hoover}
\newacronym{PR}{PR}{Parrinello-Rahman}
\newacronym{opls}{OPLS-AA}{Optimized Potentials for Liquid Simulations (all atoms)}
\newacronym{auc}{AUC}{analytical ultracentrifugation}
\newacronym{sv-auc}{SV-AUC}{sedimentation velocity AUC}
\newacronym{sfrs}{SFRS}{Soret forced Rayleigh scattering}
\newacronym{nmr}{NMR}{nuclear magnetic resonance}
\newacronym{lc}{LC}{liquid chromatography}
\newacronym{hplc}{HPLC}{high-performance liquid chromatography}
\newacronym{rplc}{RPLC}{reversed phase liquid chromatography}
\newacronym{hrtem}{HRTEM}{high resolution transmission electron microscopy}
\newacronym{stm}{STM}{scanning tunnelling microscopy}
\newacronym{xrd}{XRD}{X-ray diffraction}
\newacronym{pls}{PLS}{polystyrene latex particles}

\usepackage{latex_math_shortcuts}
\newcommand{\zeqref}[1]{\textup{(\zref{#1})}}

\begin{document}


\title{The Stokes-Einstein-Sutherland equation at the nanoscale revisited}



\author{Andreas Baer}
\affiliation{Friedrich-Alexander-Universität Erlangen-Nürnberg, Department of Physics, PULS Group, Interdisciplinary Center for Nanostructured Films (IZNF), Cauerstr. 3, 91058 Erlangen, Germany}
\author{Simon E. Wawra}
\affiliation{Friedrich-Alexander-Universität Erlangen-Nürnberg, Institute of Particle Technology (LFG), Cauerstr. 4, 91058 Erlangen, Germany}
\affiliation{Friedrich-Alexander-Universität Erlangen-Nürnberg, Interdisciplinary Center for Functional Particle Systems (FPS), Haberstr. 9a, 91058 Erlangen, Germany}
\author{Kristina Bielmeier}
\affiliation{Friedrich-Alexander-Universität Erlangen-Nürnberg, Institute of Particle Technology (LFG), Cauerstr. 4, 91058 Erlangen, Germany}
\affiliation{Friedrich-Alexander-Universität Erlangen-Nürnberg, Interdisciplinary Center for Functional Particle Systems (FPS), Haberstr. 9a, 91058 Erlangen, Germany}%
\author{Maximilian J. Uttinger}
\affiliation{Friedrich-Alexander-Universität Erlangen-Nürnberg, Institute of Particle Technology (LFG), Cauerstr. 4, 91058 Erlangen, Germany}
\affiliation{Friedrich-Alexander-Universität Erlangen-Nürnberg, Interdisciplinary Center for Functional Particle Systems (FPS), Haberstr. 9a, 91058 Erlangen, Germany}
\author{David M. Smith}
\affiliation{Ruđer Bošković Institute, Division of Physical Chemistry, Group of Computational Life Sciences, Bijenička 54, 10000 Zagreb, Croatia}
\author{Wolfgang Peukert}
\affiliation{Friedrich-Alexander-Universität Erlangen-Nürnberg, Institute of Particle Technology (LFG), Cauerstr. 4, 91058 Erlangen, Germany}
\affiliation{Friedrich-Alexander-Universität Erlangen-Nürnberg, Interdisciplinary Center for Functional Particle Systems (FPS), Haberstr. 9a, 91058 Erlangen, Germany}
\author{Johannes Walter}
\email{johannes.walter@fau.de}
\affiliation{Friedrich-Alexander-Universität Erlangen-Nürnberg, Institute of Particle Technology (LFG), Cauerstr. 4, 91058 Erlangen, Germany}
\affiliation{Friedrich-Alexander-Universität Erlangen-Nürnberg, Interdisciplinary Center for Functional Particle Systems (FPS), Haberstr. 9a, 91058 Erlangen, Germany}

\author{Ana-Sunčana Smith}
\email{smith@physik.fau.de, asmith@irb.hr}
\affiliation{Friedrich-Alexander-Universität Erlangen-Nürnberg, Department of Physics, PULS Group, Interdisciplinary Center for Nanostructured Films (IZNF), Cauerstr. 3, 91058 Erlangen, Germany}
\affiliation{Ruđer Bošković Institute, Division of Physical Chemistry, Group of Computational Life Sciences, Bijenička 54, 10000 Zagreb, Croatia}


\date{\today}

\begin{abstract}
The Stokes-Einstein-Sutherland (SES) equation is at the foundation of statistical physics, relating a particle's diffusion coefficient and size with the fluid viscosity, temperature and the boundary condition for the particle-solvent interface. It is assumed that it relies on the separation of scales between the particle and the solvent, hence it is expected to break down for diffusive transport on the molecular scale. This assumption is however challenged by a number of experimental studies showing a remarkably small, if any, violation, while simulations systematically report the opposite. To understand these discrepancies, analytical ultracentrifugation experiments are combined with molecular simulations, both performed at unprecedented accuracies, to study the transport of buckminsterfullerene C\textsubscript{60} in toluene at infinite dilution. This system is demonstrated to clearly violate the conditions of slow momentum relaxation.
Yet, through a linear response to a constant force, the SES equation can be recovered in the long time limit with no more than \SI{4}{\percent} uncertainty both in experiments and in simulations.
This nonetheless requires partial slip on the particle interface, extracted consistently from all the data. These results, thus, resolve a long-standing discussion on the validity and limits of the SES equation at the molecular scale.
\end{abstract}


\maketitle



\section{Introduction}
Diffusion was first described by Robert Brown in 1827~\cite{Brown1827}, who observed jittering of small particles in water. The fundamental framework for this phenomenon, however, emerged only a century later due to contributions of~\citet{Einstein1905},~\citet{Sutherland1905} and~\citet{Smoluchowski1906}. Firstly, the \gls{ses} equation was developed relating the diffusion coefficient to temperature and the Stokes force acting on the moving particle. Secondly, Einstein and Smoluchowski provided the keystone for the fully probabilistic formulation of diffusion. The latter was experimentally confirmed by Jean-Baptiste Perrin and his students in 1908~\cite{Perrin1908, Perrin1908a, Perrin1908b, Perrin1909}.

While very simple, easy to use and often applied for particle size determination~\cite{Miller1924, Schuck2000, Sharma2006, Alexander2013, Suess2018}, the \gls{ses} equation applies in the thermodynamic limit and requires a separation of scales between the particle and the solvent in terms of mass and size~\cite{Zwanzig1965, Espanol1993, ZoranThesis}. Therefore, it should break down for diffusive transport at the molecular scale.
Surprisingly, however, experimental studies often confirm the appropriateness of the \gls{ses} prediction for molecular diffusion, despite the clear limits imposed by the theoretical framework~\cite{Carney2011}.

The paradigmatic systems for studies of the \gls{ses} equation have been the buckminsterfullerenes due to their stability and well-defined shape. Specifically, Soret forced Rayleigh scattering was used to measure the diffusion coefficient of C\textsubscript{60} in \textit{o}-dichlorobenzene ~\cite{Matsuura2015}. They compared the radius of C\textsubscript{60}, extracted using the \gls{ses} equation, to partial molecular volume measurements~\cite{Ruelle1996}. With measurement errors of up to \SI{10}{\percent}, their measured deviation of the \gls{ses} equation was less than \SI{15}{\percent}, using slip boundary conditions on the particle-solvent interface.

C\textsubscript{60} as well as C\textsubscript{70} was used again more recently to verify the performance of the \gls{ses} equation in \gls{auc} experiments~\cite{Pearson2018}. Deviations of only about \SI{5}{\percent} were found, this time with stick boundary conditions. The reference size of fullerenes was taken from the size of the rigid carbon shell~\cite{Dresselhaus1996, Hedberg1991, Liu1991, Stephens1991, Goel2004, Murphy2014}, which notably ignore solvation effects. Furthermore, in these experiments, neither finite concentration effects of C\textsubscript{60} and C\textsubscript{70}  have been taken into account nor has the statistical significance of the results been checked. Finally, the choice of the boundary conditions was restricted to stick or slip with the conclusion that stick boundary conditions provide better results.

An alternative approach to study the limits of the \gls{ses} equation are \gls{md} simulations. They are the ideal tool for this task because they explicitly account for the atomic nature of the interactions, as well as for all the relevant time and length scales. However, unlike experiments, \gls{md} simulations systematically demonstrate discrepancies from the \gls{ses} equation for small particles~\cite{Heyes1994, Heyes1998, Walser1999, Ould-Kaddour2000, Walser2001, Ould-Kaddour2003, Schmidt2003, Schmidt2004, Li2009, ZoranThesis}. However, the accuracy of the results was often questioned due to finite system sizes~\cite{Heyes1994, Walser1999, Walser2001, Ould-Kaddour2000}, and limited sampling~\cite{Ould-Kaddour2000, Ould-Kaddour2003, Li2009} of slowly convergent power law decays~\cite{Alder1970}. Interestingly, systematic simulation studies of fullerenes have not yet been performed. 
In order to understand the apparently small deviations of the \gls{ses} equation in experiments, and large deviations in simulations, we here perform \gls{auc} measurements as well as \gls{md} simulations of C\textsubscript{60} suspended in toluene. Our aim is to combat issues of accuracy and sampling to explore the implication of different physical assumptions in modeling and measuring diffusion on the molecular scale as well as the influence of the partial molecular volume. This joint experimental and theoretical effort allows us to address the applicability of the \gls{ses} equation in the infinite dilution limit, and use the precise analysis to determine the appropriate boundary conditions at the fullerene-solvent interface. We are, therefore, able to explore advantages and limitations of each approximation used to define the \gls{ses} equation and hence provide a clear explanation of systematic discrepancy between experimental and modeling efforts appearing over the several decades.

\section{Stokes-Einstein-Sutherland equation and equilibrium statistical mechanics}\zlabel{sec:theory}

\subsection{Theoretical consideration}
In 1905, both~\citet{Einstein1905} and~\citet{Sutherland1905} published the well-known relation between the diffusion and friction coefficients $D$ and $\xi$, respectively, of a particle in a solvent
\begin{equation}\zlabel{eq:intro:einstein}
D = \frac{\kbt}{\xi}\:.
\end{equation}
Here, $T$ is the temperature and $\kB$ the Boltzmann constant. Both Einstein and Sutherland used the Stokes' formula to express the friction coefficient as a function of the fluid shear viscosity $\eta$ and the hydrodynamic radius $R_\textup{H}$ of the particle, which was assumed to be spherical:
\begin{equation}\zlabel{eq:intro:stokes}
\xi = b \pi \eta R_\textup{H}\:.
\end{equation}
The prefactor $b$ is a function of the boundary condition on the particle-solvent interface and ranges from \num{4} for a perfect slip to \num{6} for a perfect stick boundary condition, as used by Stokes'~\cite{Stokes1851} and Einstein~\cite{Einstein1905}.
Finally, by combining equations~\zeqref{eq:intro:einstein} and \zeqref{eq:intro:stokes} one obtains the \gls{ses} equation
\begin{equation}\zlabel{eq:intro:ses}
D = \frac{\kBT}{b\pi \eta R_\textup{H}}\:, 
\end{equation}
for the diffusion coefficient of a particle dispersed in a liquid. 

This approach surprisingly well relates molecular fluctuations, characterized by diffusion, with a highly coarse grained hydrodynamic friction, where molecular details are no longer resolved. Instead, they are incorporated into the boundary condition $b$.  The friction itself is defined as the ratio of a force $F$ acting on the particle and the velocity $v$ resulting from this force
\begin{equation}\zlabel{eq:intro:friction}
\xi = \frac{F}{v}\:.
\end{equation}
Already in the original derivation \cite{Einstein1905, Sutherland1905}, the force acting on the particle was presumed to arise from its collisions with the solvent molecules that in equilibrium should average to zero in the long time limit.

\begin{figure*}
    \centering
    \includegraphics[width=16.0cm]{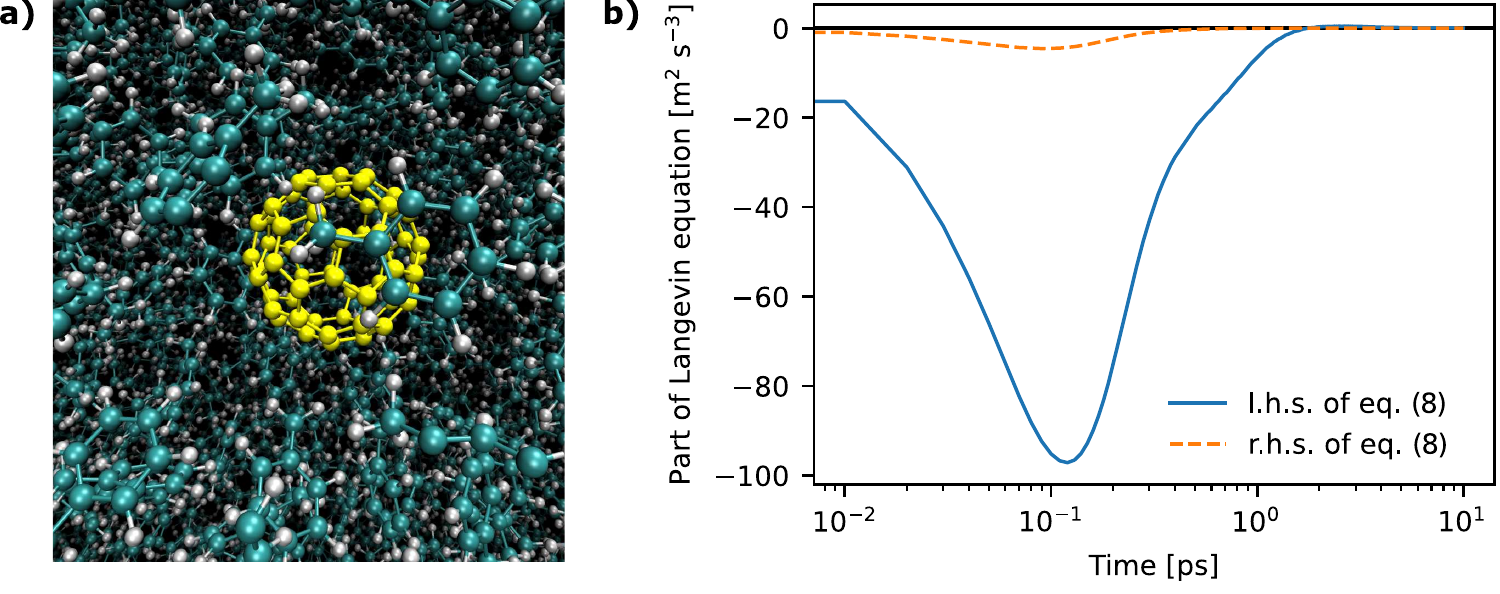}
    \caption{
    \textbf{C\textsubscript{60} in toluene as a model for studying molecular diffusion.}
    \textbf{a)} The system in question: a single C\textsubscript{60} (colored in yellow) is dispersed in toluene.
    \textbf{b)} Comparison between left and right hand side of equation~\zeqref{eq:LE-autocorr}.
    The left hand side is nearly \num{20} times larger than the right hand side. In addition, the sign is different in the regime from \SIrange{1.0}{1.7}{\ps}, demonstrating that the momentum of C\textsubscript{60} cannot be treated as a slow variable.
    Ensemble averages cover 3-dimensional trajectories of the same system size with \num{3824} solvent molecules and a total simulation time of $\approx \SI{35}{\mu\s}$.
    }
    \zlabel{fig:system}
\end{figure*}
Significant progress in understanding the relation of transport coefficients to microscopic degrees of freedom as well as the limits of applicability of the \gls{ses} equation in terms of involved time and length scales was achieved using techniques from statistical mechanics~\cite{Kirkwood1946, Green1952, Green1954, Kubo1957, Kubo1957a, Zwanzig1964, Zwanzig1965, Mori1965}.  Building on the molecular theory, it became possible to express transport coefficients as integrals of autocorrelation functions of a corresponding dynamic variable using the so called \gls{gk} relations~\cite{Kirkwood1946, Green1952, Green1954, Kubo1957a}. For example, the shear viscosity could be calculated from the off-diagonal elements of the stress tensor $P_{\alpha\beta}$
\begin{equation}\zlabel{eq:gk-visco}
\eta = \frac{V}{\kBT} \int_0^\infty \avg{P_{\alpha\beta}(t) P_{\alpha\beta}(0)} \d t\:,
\end{equation}
where the brackets $\langle . \rangle$ denote an ensemble average.
Most notably, the diffusion coefficient was found to be related to the \gls{vacf} with
\begin{equation}\zlabel{eq:gk-diff}
D_\textup{\glsentryshort{vacf}} = \int_0^\infty \avg{v(t) v(0)} \d t\:,
\end{equation}
while the friction coefficient was related to the stochastic force $F_\textup{S}(t)$ acting on the particle of interest through its \gls{acf}~\cite{Kirkwood1946, Zwanzig1964, Mori1965, ZwanzigBook}
\begin{equation}\zlabel{eq:gk-fric}
    \xi = \frac{1}{\kbt} \int_0^\infty \avg{F_\textup{S}(0) F_\textup{S}(t)} \d t\:,
\end{equation}

The latter equation can be obtained from a general Langevin equation
\begin{equation}
    \frac{\del }{\del t}p(t) = - \int_0^t \d s K(s) p(t - s) + F_\textup{S}(t) \:,
\end{equation}
where $p(t)$ is the particle momentum, $K(s)$ the memory kernel and $F_\textup{S}(t)$ a stochastic force acting on the particle.
The fluctuation-dissipation theorem relates the memory kernel to the stochastic force via
\begin{equation}
    \avg{F_\textup{S}(t) F_\textup{S}(t')} = 3 \kbt K(t - t')
\end{equation}
and the integral of the memory kernel yields equation~\zeqref{eq:gk-fric}.
Obtaining the \gls{fsacf}, required for the memory kernel, from simulations or experiments is a non-trivial task~\cite{Kowalik2019}.
However, upon deriving this formula via the projection operator formalism \cite{Mori1965, ZwanzigBook} (cf. SI section~\zref{sec:si:mori-zwanzig} for the derivation), and assuming a small rate of change of the particle momentum, i.e.
\begin{equation}
    \frac{\del}{\del t} p(t) = \ord{\lambda}
\end{equation}
with some small parameter $\lambda$, the \gls{fsacf} can be approximated with the \gls{ftacf}
\begin{equation}
    \avg{F_\textup{S}(t) F_\textup{S}(t')} = \avg{F_\textup{T}(t) F_\textup{T}(t')} + \Ord{\lambda^3}
\end{equation}
upon neglecting orders of $\lambda^3$ and higher.
The total force $F_\textup{T}$ is easily accessible through Newton's equation $F_\textup{T} = \del p/ \del t$, especially from \gls{md} simulations, where it is a prognostic variable.
The friction coefficient is then given as
\begin{equation}\zlabel{eq:gk-fric-ftacf}
    \xi = \frac{1}{\kbt} \int_0^\infty \avg{F_\textup{T}(0) F_\textup{T}(t)} \d t\:.
\end{equation}
Notably, it was shown, that the zero frequency component of the \gls{ftacf} and hence its time integral, will be non-zero if and only if the limit $\lambda \rightarrow 0$ is strictly fulfilled~\cite{Kirkwood1946, Zwanzig1964, Espanol1993, Bocquet1994, Daldrop2017}, which corresponds to a particle with constant momentum, often referred to as the frozen particle.
Nonetheless, invoking Stokes' formula~\zeqref{eq:intro:stokes} for a particle with constant, non-zero momentum also fulfills this limit.
Furthermore, both the \glspl{acf} of velocity~\zeqref{eq:gk-diff} and position (not shown here) do have a zero frequency component and can thus be used to obtain the friction coefficient~\cite{Daldrop2017}, which is equivalent to invoking the Einstein-Sutherland equation.
The statistical approach therefore provides equation~\zeqref{eq:intro:einstein} from first principles, and sets the limits of applicability of the \gls{ses} equation.

\begin{figure}
\includegraphics{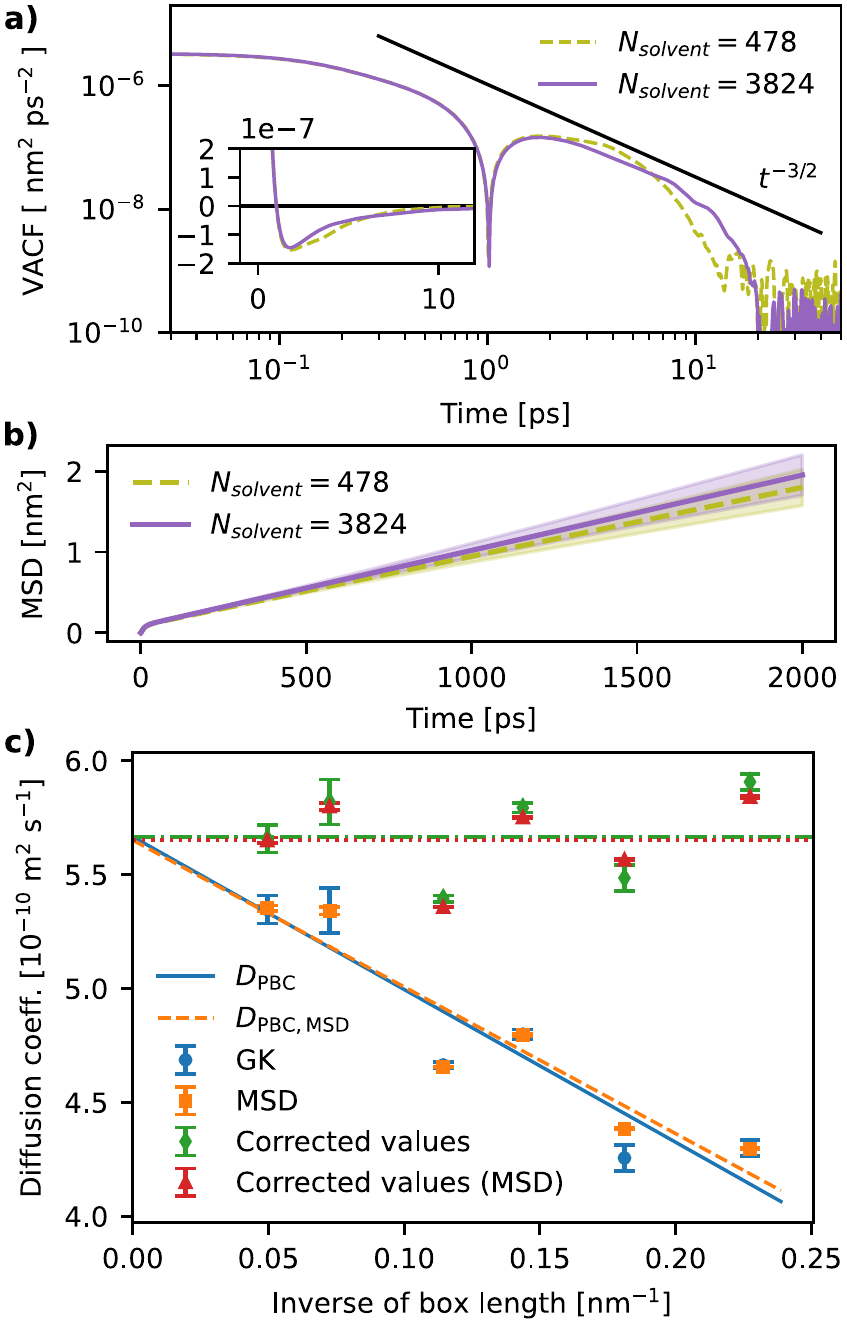}
\caption{\textbf{Diffusion coefficient of C\textsubscript{60} in toluene obtained from \gls{md} simulations.}
  \textbf{a)}
  Log-log representation of the absolute value of the \glsentryshort{vacf} for two different system sizes (inverse of box length: $\approx \SI{0.23}{\per\nm}$ and \SI{0.11}{\per\nm}).
  The analysis runs over \SI{6}{\micro\s} and \SI{35}{\micro\s} 3-dimensional trajectory, for the small and large systems, respectively.
  Especially for the larger system, the expected decay proportional to $t^{-3/2}$ is apparent for times $>\SI{2}{\ps}$~\cite{Alder1970}.
  The inset shows a lin-lin representation of the \glsentryshort{vacf} in the region around the first minimum.
  \textbf{b)}
  \glsentryshort{msd} for the same system sizes as in a).
  Due to finite size effects, the larger system produces a larger \glsentryshort{msd} and diffusion coefficient (cf. panel~c).
  \textbf{c)}
  Size dependence of the diffusion coefficient.
  Equation~\zeqref{eq:diff-size_dependence} is used for both the fit (lines) and the corrected values.
  Symbols represent mean and standard deviation of the plateau region for a single system size each (cf. SI-Figure~\zref{fig:si:diff}). The analysis runs over \SI{6}{\micro\s} to \SI{35}{\micro\s} 3-dimensional trajectory, for the different systems.
  }
\zlabel{fig:results:diffusion}
\end{figure}

\subsection{Molecular dynamics simulations}\zlabel{sec:md}
%
To assess how crucial the restriction of the constant particle momentum is,
we reformulate the problem  (cf. SI section~\zref{sec:si:mori-zwanzig} for the derivation), to obtain a Volterra equation of first kind, that can be directly checked from simulations:
\begin{equation}\zlabel{eq:LE-autocorr}
\frac{\del }{\del t}\avg{p(t) p(0)} = -\frac{1}{\kbt} \int_0^t \d s \avg{F_\textup{T}(s) F_\textup{T}(0)} \cdot \avg{p(t) p(0)}\:.
\end{equation}
This equation includes approximating the \gls{fsacf} with the \gls{ftacf}.

\Gls{md} simulations are performed using the GROMACS simulation package~\cite{Gromacs1, Gromacs2, Gromacs3, Gromacs4, Gromacs5, Gromacs6, Gromacs7}, by placing a single C\textsubscript{60} in a box of toluene (cf. Figure~\zref{fig:system}a) and applying periodic boundary conditions (see SI section~\zref{sec:si:methods} for details). For toluene, we use the \gls{opls} force field~\cite{OPLSAA1}, while the parameters for C\textsubscript{60} are taken from previous work \cite{Baer2022}. Following an extensive protocol to equilibrate the system at room temperature and \SI{1}{\bar}, production simulations are performed in NVT ensemble with the temperature maintained by the Nosé-Hoover thermostat. We modify the output routine of GROMACS to simultaneously extract the total force on the particle and its momentum with the output rate of \SI{10}{\fs} without all solvent information.
This accounts for C\textsubscript{60} internal forces and external van der Waals forces, while Coulomb forces are absent for the uncharged C\textsubscript{60} atoms.
To take into account finite size effects we simulate an array of systems with 478 to 46838 toluene molecules (Figure~\zref{fig:system}a), for in total \SIrange{0.6}{35}{\micro\s}, and apply the appropriate finite size corrections \cite{Duenweg1993, Yeh2004}. Using the system isotropy, averaging over all spatial dimensions is performed to improve the statistics.

This methodology allows us to evaluate both sides of equation~\zeqref{eq:LE-autocorr} by numerical differentiation or integration. We find that this expression does not hold for a C\textsubscript{60} in toluene, which clearly demonstrates that the momentum is not a slowly changing variable (Figure~\zref{fig:system}b). Thus, one could infer that the friction coefficient cannot be properly obtained by the \gls{gk} approach, which becomes evident from directly evaluating equation~\zeqref{eq:gk-fric-ftacf}.
The full integral vanishes as expected \cite{Kirkwood1946, Daldrop2017}.
Other methods were previously suggested to calculate the friction coefficient from the running integral of the \gls{ftacf} \cite{Espanol1993, Ould-Kaddour2003, ZoranThesis}, which worked for particles heavy compared to the solvent molecules but not necessarily infinitely heavy.
In our case, neither a linear nor an exponential decay can be observed (cf. SI section~\zref{sec:si:diffusion:facf} and SI-Figure~\zref{fig:results:facf}) and thus the only of these methods left is using the maximum of the running integral, that corresponds to the first zero-crossing of the \gls{ftacf}, as suggested by \citet{Lagarkov1978}.
This method, known to be an overestimation~\cite{Espanol1993, Ould-Kaddour2003}, yields with corrections for finite size effects (see SI section~\zref{sec:si:finite-size} and \zeqref{eq:friction-size}) $\xi = \SI{3.37+-0.03e-12}{\kg\per\s}$.


This can be verified by extracting the diffusion coefficient of C\textsubscript{60} from  the \gls{vacf} via the \gls{gk} relation \zeqref{eq:gk-diff} and the \gls{msd} (cf. SI section~\zref{sec:si:diffusion:vacf}) as both methods are well established~\cite{Alder1970, ZoranThesis} (cf. Figure~\zref{fig:results:diffusion}a,b).
We account for the finite size effects by the theoretical size dependence of the diffusion coefficient \cite{Duenweg1993, Yeh2004}, adjusted to include a variable boundary condition $b$
\begin{equation}\zlabel{eq:diff-size_dependence}
D_\infty = D_\textup{PBC} + \frac{\kBT \zeta}{b\pi\eta a}\:.
\end{equation}
Using this equation, all data (cf. Figure~\zref{fig:results:diffusion}c) can be fitted (blue circles and orange squares) or corrected (green diamonds and red triangles). Hereby, the fit was calculated by treating both the boundary condition parameter $b$ and the infinite diffusivity $D_\infty$ as free parameters, yielding $D_\infty = \SI{5.67 \pm 0.19 e-10}{\m\squared\per\s}$ for the \gls{vacf} and  $D_\infty = \SI{5.65 \pm 0.17 e-10}{\m\squared\per\s}$ for the \gls{msd}. Unfortunately, the parameter $b$ carries a significant uncertainty, such that precise conclusions about the effective boundary condition cannot be made using this approach. 

Using equation \zeqref{eq:intro:einstein}, we compare the obtained $D_\infty$, with the calculated $\kbt/\xi$. We find deviations larger than \SI{50}{\percent}, due to the gross underestimation of the C\textsubscript{60} friction coefficient. This clearly shows that the standard approximations of equilibrium statistical physics underlying the \gls{ses} equation do not hold in this system, presumably because of the similar mass of the C\textsubscript{60} and the toluene molecule.

As this deviation stems from the failure of approximating the memory kernel $K(s)$ with the \gls{ftacf} instead of the \gls{fsacf}, it is only natural to verify whether the memory kernel, if extracted with different methods~\cite{Kowalik2019}, reproduces the correct friction coefficient.
Several methods to derive the memory kernel for systems on the molecular scale have been developed in the past years~\cite{Kowalik2019}.
We calculate $K(s)$ through the Fourier transform from the \gls{vacf}, as suggested by~\cite{Gottwald2015, Kowalik2019}, which is equivalent to calculating the diffusion coefficient from the \gls{vacf} and invoking the Einstein-Sutherland equation~\zeqref{eq:intro:einstein}.
For the system with 3824 solvent molecules, we obtain a clear deviation from the kernel obtained from the \gls{ftacf} (see SI-Figure~\zref{fig:results:memory-kernel}a).
By integrating the kernel from the \gls{vacf}, we obtain $\xi_\textup{PBC} = \SI{8.68+-0.03 e-12}{\kg\per\s}$, which, by the Einstein-Sutherland equation~\zeqref{eq:intro:einstein}, corresponds to a diffusion coefficient of $D_\textup{PBC} = \SI{4.66+-0.02 e-10}{\m\squared\per\s}$, precisely the value obtained by integrating the \gls{vacf} (both values are for the same system size and thus not corrected for system size effects).
This expected agreement underlines the failure of approximating the \gls{fsacf} with the \gls{ftacf} for such small particles, while supporting the validity of the Einstein-Sutherland equation.

\section{Einstein-Sutherland equation under conditions of constant drag}
\subsection{Analytical ultracentrifugation experiments}\zlabel{sec:AUC}
\begin{figure}
\centering
\includegraphics{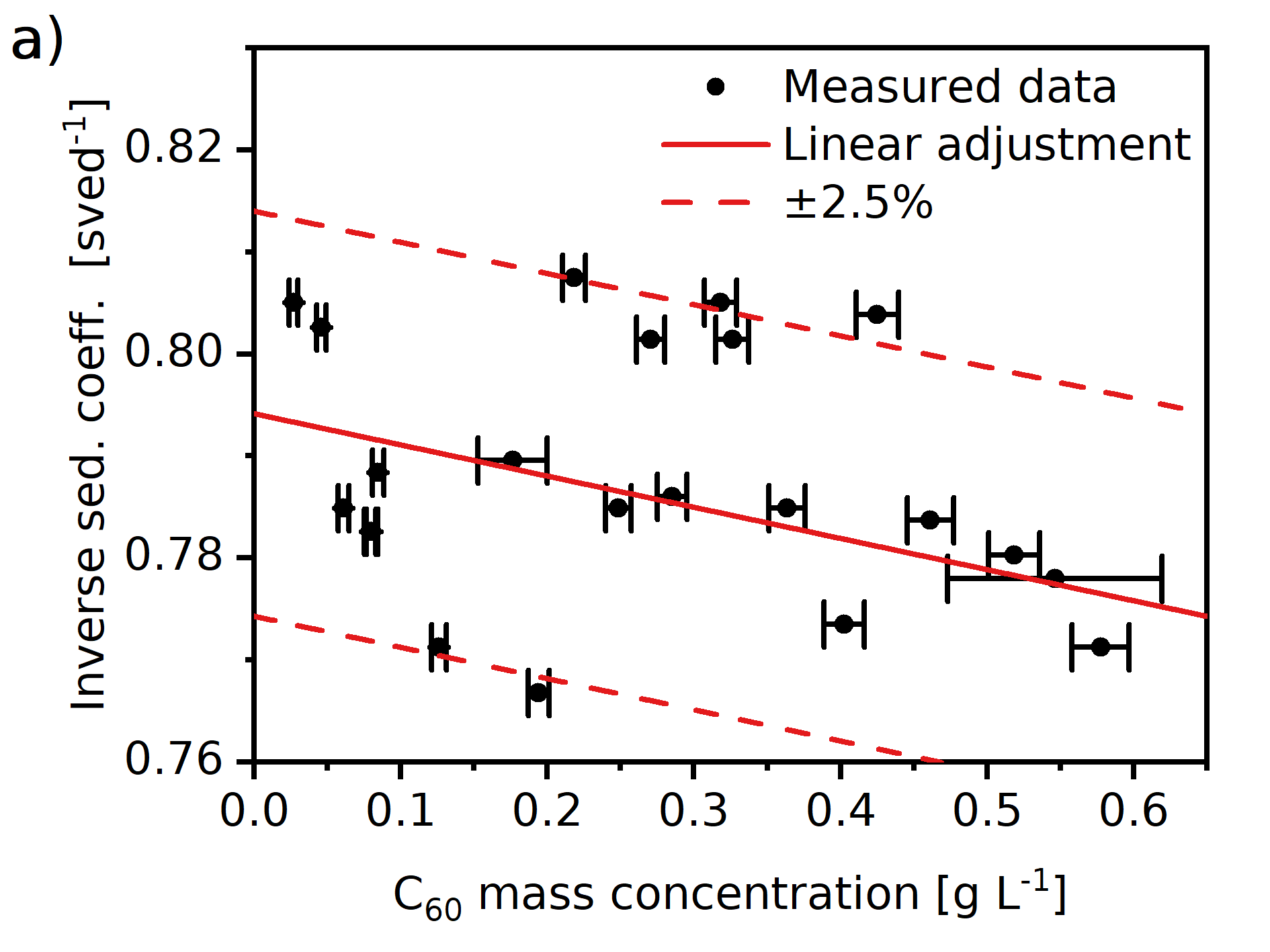}
\includegraphics{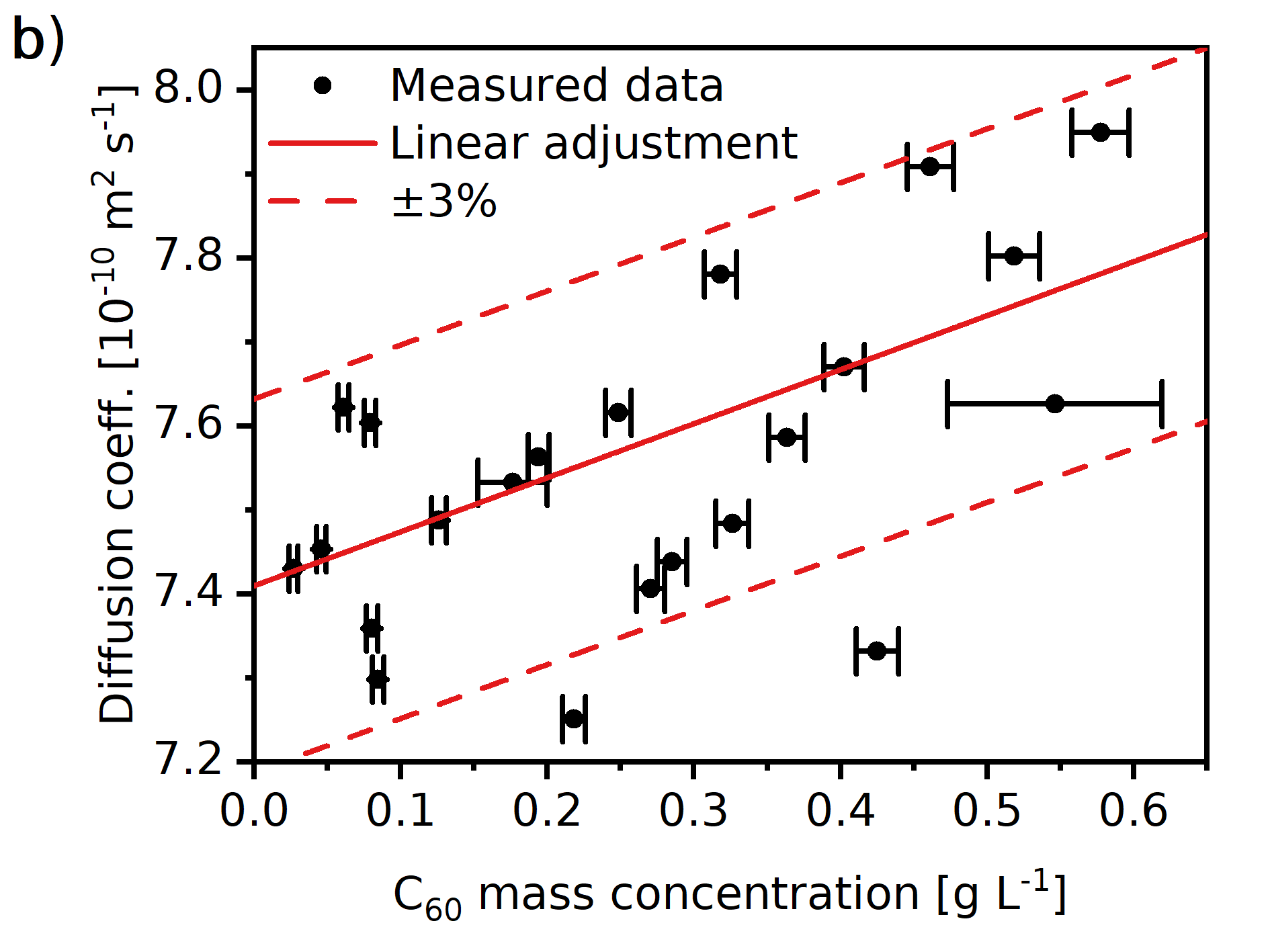}
\caption{\textbf{Retrieved inverse sedimentation coefficients (a) and diffusion coefficients (b) from \glsentryshort{sv-auc} experiments as a function of C\textsubscript{60} mass concentration.}
Depicted are hydro- and thermodynamic non-ideality of C\textsubscript{60} resulting from data-analysis with the $c(s)$-model (Levenberg-Marquardt). Linear fits were performed taking the uncertainty of the extinction, optical path length and the extinction coefficient at \SI{570}{\nm} for the concentrations into account. The corresponding uncertainty (standard deviation) is depicted with the symbols.}
\zlabel{fig:results:AUC}
\end{figure}
There are several reports that validate the \gls{ses} equation for C\textsubscript{60} in molecular solvents~\cite{Carney2011, Matsuura2015, Pearson2018}. However, even these works differ in the choice of the boundary conditions and treatment of effects of finite concentration. This, together with the results of the previous section, calls for a discussion of the validity of the theory and/or the correct choice of parameters for comparison.

To clarify these issues, we perform a set of \gls{auc} experiments to observe the reaction of the system to a centrifugal field~\cite{Walter2014, Schuck2016, Uchiyama2016}.
In this type of experiments, referred to as \gls{sv-auc} experiments, we resolve the temporal evolution and radial distribution of the particles’ concentration (cf. SI section~\zref{sec:si:auc-experiments}).
Using numerical solutions of the Lamm equation \zeqref{eq:Lamm}, we fit the data with the diffusion and sedimentation coefficients $D$ and $s$, respectively, while also considering compressibility of the solvent (cf. SI sections~\zref{sec:si:auc-formulas}, \zref{sec:si:auc-interpretation} and~\zref{sec:si:diffusion:auc} for details).
Conducting the experiment at different concentrations (Figure~\zref{fig:results:AUC}), allows us to retrieve the values extrapolated to infinite dilution by means of a linear fit to be $D_0 = \SI{7.41 +- 0.04}{\m\squared\per\s}$ and $s_0 = \SI{1.26 \pm 0.01}{sved}$.
The single sample measurement of \citet{Pearson2018} at a finite concentration, which we estimate to be about \SI{1}{\g\per\l}, yields $D_\textup{Pearson} = \SI{7.59}{\m\squared\per\s}$, which is fully consistent with our data.

However, direct comparison of the measured diffusion coefficients to simulation results (section~\zref{sec:md}) shows significant differences.
This discrepancy can be attributed to the \num{1.4} times higher viscosity of toluene provided by the \gls{opls} force field~\cite{OPLSAA1}  (see SI section~\zref{sec:si:diffusion:viscosity}) compared to experimental reference measurements at same thermodynamic conditions of a temperature of \SI{293.15}{\K} and pressure of \SI{1}{\bar} \cite{Santos2006}.
We can, nonetheless, normalize the diffusion coefficient by the ratio of the \gls{md} and measured viscosity (SI section~\zref{sec:si:diffusion:viscosity}), which provides an agreement between the simulated and the experimental diffusivity with a deviation of less than \SI{10}{\percent} (see Table~\zref{tab:results:summary}, column 4).

We can then obtain the friction coefficient directly from $s_0 = m / \xi$, with $m$ being the excess mass of the analyte, i.e. the mass of the analyte minus the mass of solvent with the same volume.
\citet{Ruelle1996} measured the C\textsubscript{60} partial molecular volume in toluene to be \SI{0.603}{\nm\cubed}, which combined with
its molecular mass of \SI{720.66}{\atomicmassunit} and a toluene density of \SI{866.86}{\kg\per\m\cubed} yields a friction coefficient of \SI{5.35e-12}{\kg\per\s}.
We can now compare this to the prediction of the Einstein-Sutherland equation~\zeqref{eq:intro:einstein} obtained from the diffusion coefficient of the \gls{auc} experiments, which gives $\kbt / D_0 = \SI{5.46e-12}{\kg\per\s}$.
Both results agree within \SI{2}{\percent},
clearly demonstrating the validity of the Einstein-Sutherland equation for this system, which was not explicitly shown before.
We hypothesize that this estimate of friction stems from the non-equilibrium nature of the \gls{auc} experiment. Namely rather than the friction being measured as a response to the stochastic force, here it emerges from the response to a constant drag or centrifugal force acting on the C\textsubscript{60}.


\subsection{Friction as response to a drag force in non-equilibrium MD simulations}\zlabel{sec:diffusion:pull}
\begin{figure}
\centering
\includegraphics{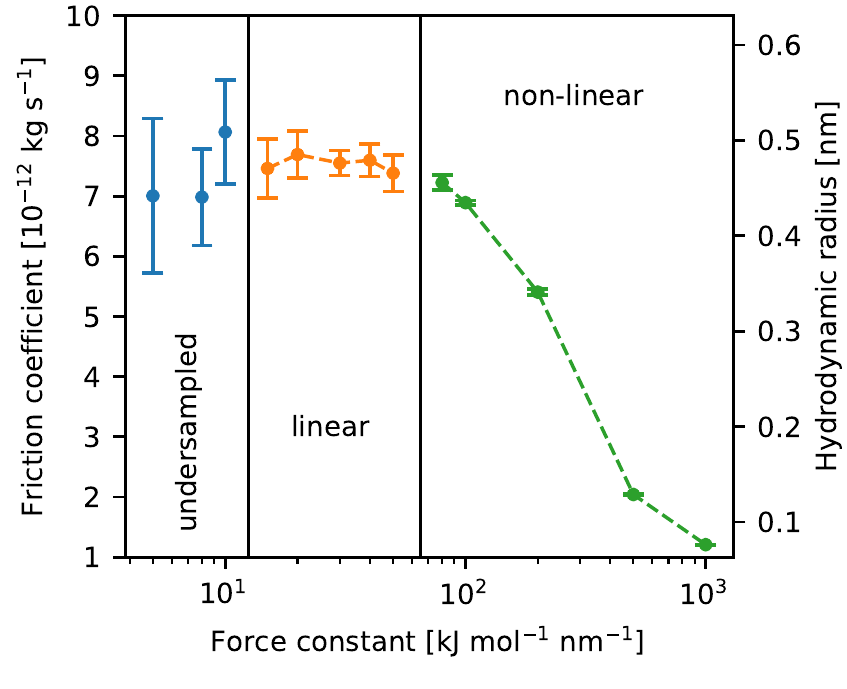}
\caption{\textbf{Friction coefficient of C\textsubscript{60} obtained from steered \glsentryshort{md} simulations.}
The friction coefficient $\xi_\textup{PBC}$ is calculated for each applied constant force as the ratio of applied force and resulting average velocity.
Three regimes are indicated, the regime where statistics are not sufficient to determine a precise friction coefficient, the regime of proper linear response and the regime of non-linear response, where the apparent friction coefficient drops significantly.
For the present study, only the linear regime is relevant and used to determine the friction coefficient as the mean over the entire range.
Displayed error bars denote the uncertainty in the mean of the calculated friction coefficient, where the mean is calculated over a total of \SIrange{40}{200}{\ns} trajectory.
The hydrodynamic radius displayed corresponds to a boundary condition value of $c=6$.}
\zlabel{fig:results:pull-friction}
\end{figure}
To verify our hypothesis that the friction coefficient can be determined if a particle is subject to a non-vanishing average force, we return to modeling and perform non-equilibrium molecular dynamics NEMD simulations. Specifically, an additional (constant) force on the fullerene is added while removing the center-of-mass motion of the entire system. The latter is required to obtain a proper frame of reference with periodic boundary conditions. The resulting particle velocity, relative to the fluid, is then calculated and combined with the known force to extract the friction coefficient $\xi$ from equation~\zeqref{eq:intro:friction}.

To ensure, that we sample the linear regime, a set of pull forces is investigated (cf. Figure~\zref{fig:results:pull-friction} and SI section~\zref{sec:si:diffusion:pull}).
For weak pull forces (blue points in Figure~\zref{fig:results:pull-friction}), the sampling is poor due to the small signal-to-noise ratio, while for large pull forces (green points), non-linear effects arise.
For intermediate pull forces (orange points), we observe a properly converged and linear regime. Accounting for finite size effects \zeqref{eq:friction-size} yields $\xi = \SI{6.9+-0.15 e-12}{\kg\per\s}$.
Notably, this again compares well (about \SI{10}{\percent} deviation) to the \gls{auc} result, when corrected for the differences in the viscosity (cf. section~\zref{sec:AUC}).

Importantly, we can calculate the diffusion coefficient from this friction coefficient with the Einstein-Sutherland equation~\zeqref{eq:intro:einstein} to obtain $D = \SI{5.9 +- 0.1 e-10}{\m\squared\per\s}$, with the statistical error of \SI{2}{\percent}. This is basically the same result as the diffusion coefficient obtained from the \gls{msd} or the \gls{vacf} (section~\zref{sec:md}).
This confirms the validity of the Einstein-Sutherland equation found in the \gls{auc} experiments also for \gls{md} simulations, rendering both techniques consistent.

\section{Stokes-Einstein-Sutherland equation}
\subsection{The effective radius of C\textsubscript{60}}\zlabel{sec:radius}
\begin{table}
\begin{ruledtabular}
\caption{\textbf{Summary of radii of the C\textsubscript{60} obtained by various methods.}
The radius $R_\textup{graphite}^\textup{{exp}}$ is obtained by adding half of the graphite interplanar distance to the radius of the structure of the C\textsubscript{60} nuclei, which gives the radius in the gas phase.
The radius $R_\textup{cryst}^\textup{{exp}}$ is obtained by the fullerene distance in C\textsubscript{60} crystal.
All radii $R_V$ are volume equivalent radii, i.e. radii corresponding to the partial molecular volume or the \gls{psv}.
}
\begin{tabular}{llcccc}
    Method & source & radius [\si{\nm}]  \\
    \hline
  $R_\textup{graphite}^\textup{{exp}}$  %
    &~\cite{Yannoni1991, Johnson1992, Hedberg1991, Liu1991, Heiney1991, Stephens1991} %
    & \num{0.522 +- 0.001}\footnotemark[1]&  \\
    $R_\textup{cryst}^\textup{{exp}}$%
    &~\cite{Kraetschmer1990, Wilson1990, Goel2004} %
    & \num{0.52 +- 0.01}\footnotemark[1]& \\
    $R_{\textup{V}}^{\textup{sim}}$ %
    & section~\zref{sec:radius} %
    & \num{0.500 +- 0.003}& \\
    
    $R_{\textup{V}}^{\textup{exp, AUC}}$ %
    & section~\zref{sec:radius} %
    & \num{0.520 +- 0.002}& \\
    
    $R_{\textup{V}}^{\textup{exp, DC SE}}$ %
    & section~\zref{sec:si:radius} %
    & \num{0.526 +- 0.014}&  \\
    
    $R_{\textup{V}}^{\textup{exp}}$ %
    & \cite{Ruelle1996} %
    & \num{0.524 +- 0.003}&  \\
\end{tabular}
\zlabel{tab:radii:summary}
\end{ruledtabular}
\footnotetext[1]{This value is the average over several measurements reported in the literature (see SI section~\zref{sec:si:radius}).}
\end{table}

With the validity of the Einstein-Sutherland equation demonstrated for both experiments and NEMD simulations, the next step is to assess the validity of the \gls{ses} equation and more importantly to retrieve a proper value for the boundary condition coefficient $b$.
To assess this, we need to calculate and/or measure toluene viscosity (section~\zref{sec:si:diffusion:viscosity}), and the hydrodynamic radius of the C\textsubscript{60} independently from the friction and diffusion coefficients.

As the C\textsubscript{60} is a very rigid, nearly spherical molecule that does not feature extraordinarily strong interactions with the carbon-rich toluene, solvation effects in this specific environment were found to be very small compared to the C\textsubscript{60} in the solid phase \cite{Ruelle1996}. This is reflected in sub-angstrom differences in radii $R_V$, that can be associated to a sphere with equivalent volume  (cf. \cite{Yannoni1991, Johnson1992, Hedberg1991, Liu1991, Heiney1991, Stephens1991, Kraetschmer1990, Wilson1990, Goel2004, Ruelle1996}, table~\zref{tab:radii:summary} and SI section~\zref{sec:si:radius}). Because of these small differences, $R_V$ has been used as the estimate for the C\textsubscript{60} hydrodynamic radius $R_H$ \cite{Pearson2018}.

To verify these results and validate our own simulation and experimental protocols, we pursue the independent calculation of the partial molecular volume. In simulations, the partial molecular volume can be obtained from the particle-solvent \gls{rdf}  $g(r)$ via a Kirkwood-Buff integral~\cite{Koga2013}
\begin{equation}\zlabel{eq:radius-kb}
V =\frac{4}{3}\pi R_\textup{V}^3= \frac{1}{n_\textup{s}} + 4\pi \int_0^\infty [g_\textup{s}(r) - g(r)]r^2 \d r\:,
\end{equation}
with $g_\textup{s}(r)$ being the \gls{rdf} of the pure solvent and $n_\textup{s}$ the number density of the solvent (cf. SI-Figure~\zref{fig:results:psv}a).
This provides $R_\textup{V}^\textup{{sim}} = \SI{0.500+-0.003}{\nm}$, deviating from the experimental reference \cite{Ruelle1996} by only \SI{4}{\percent}.

We, furthermore, independently determine the C\textsubscript{60} radius in experiments, using sedimentation equilibrium \gls{auc} (SE-AUC) experiments with density contrast.
In these measurements, the rotor speed and thus centrifugal force is sufficiently small such that the sedimentation flux compensates the diffusion flux at each radial position.
From the resulting exponential concentration profile, the apparent buoyant mass can be calculated independent of the viscosity. Using  solvents with different levels of deuteration, which changes solvent density but not the solvation properties of dissolved species, the \gls{psv} can be obtained. Due to a small number of parameters and thus small statistical uncertainties, the relevant data can be obtained with high accuracy yielding $R_\textup{V}^\textup{exp, DC} = \SI{0.526+-0.014}{\nm}$, which is within measurement errors of less than \SI{3}{\percent} compared to literature results obtained using different methods~\cite{Ruelle1996}, and only \SI{5}{\percent} different to the simulation data. Finally to check the consistency of our \gls{sv-auc} experiments, we retrieve the \gls{psv} directly from the diffusion and sedimentation coefficients at the infinite dilution limit by rearranging the well-known Svedberg equation~\zeqref{eq:Svedberg} to solve for the \gls{psv} \zeqref{eq:psv-auc}.
This gives $R_{\textup{V}}^{\textup{exp, AUC}} = \SI{0.520 +- 0.002}{\nm}$.
Notably, this is within measurement errors of less than \SI{1}{\percent} equivalent to the values obtained with our other approaches and the literature \cite{Ruelle1996} (cf. Table~\zref{tab:radii:summary}).

\subsection{Effective boundary condition at the C\textsubscript{60} toluene interface}\zlabel{sec:boundary-condition}
\begin{table*}
\begin{ruledtabular}
\sisetup{
    propagate-math-font = true ,
    reset-math-version = false
}
\caption{\textbf{Summary of diffusion coefficients and boundary conditions of the C\textsubscript{60} obtained by various methods.}
Values derived using \zeqref{eq:intro:einstein} are indicated by superscripts.
All values except for the one obtained from the F\textsubscript{T}ACF match.
We find the averaged value of $\bar{b}=\num{5.5+-0.2}$ (cf. section~\zref{sec:boundary-condition}) to be consistent with all other presented data.
}
\begin{tabular}{llcccc}
    Method & source  & diffusivity & rescaled diffusivity & $b$ from \gls{ses} \\
    &  & [\SI{e-10}{\m\squared\per\s}] & [\SI{e-10}{\m\squared\per\s}] & [-] \\
    \hline
    F\textsubscript{T}ACF
    & section~\zref{sec:si:diffusion:facf}%
    & \num{12.0 +- 0.1}\footnotemark[1]%
    & \num{17.4 +- 0.2}\footnotemark[1]%
    & \num{2.53 +- 0.04} \\ 
    VACF, MSD
    & section~\zref{sec:md}%
    & \num{5.7 \pm 0.2}%
    & \num{8.2 \pm 0.3}%
    & \num{5.3 +- 0.2} \\ 
    Pulling force
    & section~\zref{sec:diffusion:pull}%
    & \num{5.9 +- 0.1}\footnotemark[1]%
    & \num{8.5 +- 0.2}\footnotemark[1]%
    & \num{5.2 +- 0.2} \\ 
    AUC exp. %
    & section~\zref{sec:si:diffusion:auc} %
    &  \num{7.41 +- 0.04}%
    &  \num{7.41 +- 0.04}%
    & \num{5.71 +- 0.06} \\ 
    AUC exp. %
    &~\cite{Pearson2018}%
    &  \num{7.59}%
    &  \num{7.59}%
    & \num{5.54 +- 0.03}\footnotemark[2] \\ 
\end{tabular}
\zlabel{tab:results:summary}
\end{ruledtabular}
\footnotetext[1]{This value is calculated using equation~\zeqref{eq:intro:einstein}.}
\footnotetext[2]{This value is derived in conjunction with the radius obtained from \citet{Ruelle1996}.}
\end{table*}

With these referent values for the C\textsubscript{60} size, we can now proceed with the determination of the boundary condition at the particle-solvent interface. For the \gls{auc} experiments, the boundary condition value can be retrieved directly from the well-known frictional ratio $\xi/\xi_0$, which is also known as $f/f_0$ in the \gls{auc} literature. It is defined as the ratio of the friction coefficient of the analyte to the friction coefficient of a sphere of equal volume as the analyte and assuming stick boundary conditions.
When expressed in terms of $s$ and $D$ and using the \gls{psv} $\bar{\nu}$ of the fullerene, one obtains for spherical particles the well-known form:
\begin{equation}
\zlabel{eq:frictional_ratio}
    \xi/\xi_0 = \left(\frac{\sqrt{2}}{18\pi} \frac{\kbt}{D \sqrt{s} \eta^{\frac{3}{2}}} \sqrt{\frac{1-\bar{\nu}\rho}{\bar{\nu}}}\right)^{\frac{2}{3}} = \num{0.95+-0.01}
\end{equation}
The frictional ratio is typically used to evaluate shape anisotropy and volume expansion due to solvation, when $\xi/\xi_0\geq 1$~\cite{Schuck2016}.
With the nearly perfect spherical shape of C\textsubscript{60} and hardly any volume expansion effects (see SI sections~\zref{sec:si:radius:static} and~\zref{sec:si:radius}), we expect a value very close to unity.
However, values $\xi/\xi_0 < 1$, on the other hand, point to deviations from the stick boundary condition. Indeed, when assuming $R_\textup{H} = R_\textup{V}$, we obtain $b = 6 \cdot \xi/\xi_0 = \num{5.71+-0.06}$.

We can furthermore determine the boundary condition $b$ at the C\textsubscript{60}--toluene interface, such that the \gls{ses} equation holds, also from simulations. Using $D_\textup{VACF}$ or $D_\textup{MSD}$ and equation~\zeqref{eq:intro:ses}, with $R_\textup{H} = R_\textup{V}$ 
determined in simulations, we obtain $b = \num{5.3+-0.2}$.
Following a similar strategy, we can combine experimental data in the literature ($R_\textup{V}^\textup{exp}$~\cite{Ruelle1996} and $D_\textup{AUC}$~\cite{Pearson2018}), and obtain $b = \num{5.54+-0.03}$, which is within the error bar of the simulation and only \SI{3}{\percent} smaller than our experimental results.

Due to significant accuracy of the measurements and simulations, these results clearly indicate that perfect stick boundary conditions, typically assumed in experiments \cite{Pearson2018, Carney2011, Li2003, Tuteja2007, Longsworth1952, Cheng1955, Feitosa1991} and simulations \cite{Heyes1994, Heyes1998, Schmidt2003, Schmidt2004, Li2009}, may not be the correct choice for C\textsubscript{60}. Actually, with $b=6$ systematic errors of \SIrange{5}{12}{\percent} are found in simulations and experiments. Furthermore, $R_\textup{H}$ calculated from the \gls{ses} is found to be \emph{smaller} than $R_\textup{V}$. 

Finding deviations from the stick boundary conditions should not be surprising in the light of a long going discussion on its application for small particles~\cite{Cunningham1910}.
The argument is captured by the Knudsen number $\mathit{Kn}$, the latter being defined as the ratio of mean free path and size of a particle. It can be calculated as $\mathit{Kn} = ( \sqrt{2}\pi n d^2 l )^{-1}$, where $n$ and $d$ are number density and diameter of the solvent, while $l$ is the characteristic length of the system.
With $d$ taken to be the diameter of the toluene aromatic ring and $l$ the diameter of the fullerene, we estimate $\mathit{Kn} \approx \num{0.09}$. This is significant because the Knudsen number can be used as a measure to determine $b$~\cite{Li2003}.
Specifically for $\mathit{Kn} = 0$, one expects perfect stick ($b=6$) while for $\mathit{Kn} \rightarrow \infty$, perfect slip ($b=4$). 
As soon as $\mathit{Kn}$ is not vanishing, as in the present case, $b<6$ should be obtained, which is indeed the case.

\section{Discussion and Conclusions}\zlabel{sec:conclusions}
We here present a set of experimental and simulation results on the dynamic and static properties of a C\textsubscript{60} dispersed in toluene. We perform \gls{auc} experiments reporting diffusion and sedimentation coefficients in the infinite dilution limit, improving on the accuracy of the method \cite{Pearson2018}. Instead of following the usual approach to \gls{auc} data, and calculating the particle mass and size, we use the known mass of C\textsubscript{60} to report the hydrodynamic radius and the boundary condition at the particle-solvent interface. 
We furthermore perform a quantitative comparison of simulations and experiments which is excellent for static properties derived from density distributions. This includes the determination of the partial molecular volume of C\textsubscript{60}, which differs from the experimental estimate by only a couple of percent. Extracting dynamic properties such as viscosity is more challenging due to the limitation of current force fields. However, upon simple re-scaling by the viscosity contrast (column~4 in Table~\zref{tab:results:summary}) the  difference between observed and measured diffusivities is only \SI{10}{\percent}.
Furthermore, the analysis which does not rely on correcting for the viscosity contrast, allows us to make several important findings:
\begin{itemize}
    \item The expression associating the \gls{facf} and the friction coefficient suggested by~\citet{Kirkwood1946}, the Green-Kubo theory~\cite{Green1952, Green1954, Kubo1957, Kubo1957a} and the Mori-Zwanzig formalism~\cite{Zwanzig1964, Zwanzig1965, Mori1965} is the main starting point for the critique on the applicability of the \gls{ses} equation at the nanoscale. In its derivation, the assumption that the particle momentum is constant, or only very slowly changing, is mandatory. We show that this assumption is clearly violated for C\textsubscript{60} in toluene due to its small molecular weight ~\cite{Espanol1993,Bocquet1994}, back-scattering and dissipation of momentum through internal degrees of freedom that couple with directional motions on sub-picosecond time scales (Figure~\zref{fig:system}b). The \gls{ftacf} integrates these effects, resulting in a vanishing response of the zero frequency component and thus, in this formulation, cannot be directly related to friction. If, contrastingly, the memory kernel is obtained from the \gls{fsacf} or via a different method \cite{Kowalik2019, Gottwald2015}, the actual friction coefficient can be obtained.
    
    \item Another reliable measurement of the friction coefficient, however, can be obtained in non-equilibrium conditions. The good estimate can be extracted from the average velocity of C\textsubscript{60} induced by a drag force in the linear response. This is permitted by the momentum conservation and the nanosecond sampling times when conditions of slow dynamics are recovered (Figure~\zref{fig:system}b).  In experiments, the non-equilibrium conditions are provided in the sedimentation experiments, while in simulations, the friction is obtained in steered molecular dynamics (following equation~\zeqref{eq:intro:friction}). Using the Einstein-Sutherland relation, we can compare this friction to independently measured diffusion constants.  For experiments, where infinite dilution sedimentation and diffusion coefficients are extracted by a sequence of measurements, the Einstein-Sutherland equation is recovered with \SI{1}{\percent} accuracy. In simulations, by accounting for finite size effects, the Einstein-Sutherland relation holds with a precision of \SI{<2}{\percent}. This confirms the validity of the Einstein-Sutherland relation ~\zeqref{eq:intro:einstein} at the nanoscale for long observation times.
    
    \item Under the assumption of stick ($b=6$), the hydrodynamic radius, as measured from diffusion data or from the response to drag is systematically smaller than the radius calculated directly from the partial molecular volume associated with the second virial coefficient. With this assumption, we can also verify the validity of the \gls{ses} equation, which is found with $10-15\%$ precision.
    
    \item Using the size of the particle obtained from the partial molecular volume, the independently obtained friction coefficient and viscosity, we deduce the boundary condition on the particle with equation~\zeqref{eq:intro:stokes}.  Averaging over all experimental and simulation data, we find small deviations from perfect stick ($\bar{b}=\num{5.5+-0.2}$). This is fully consistent with the Knudsen number for C\textsubscript{60} in toluene.
    
    \item Using partial slip, all experimental and simulation data become consistent with the Stokes-Einstein-Sutherland equation~\zeqref{eq:intro:ses}. Acquired potential errors are within the statistical uncertainties of \SI{2}{\percent} to \SI{4}{\percent} (cf. Table~\zref{tab:results:summary}). This demonstrates a quantitative agreement of simulation results and carefully acquired experimental data on the validity of the \gls{ses} equation, in a real system on the 1 nm length scale, and away from the infinite mass limit of the solute, which was a decades old problem.  
\end{itemize}
    
In conclusion, our findings are instrumental to explain the reason for small or no inconsistencies of the Stokes-Einstein-Sutherland equation on the nanoscale~\cite{Carney2011, Jin2010, Matsuura2015, Pearson2018}, despite violation of basic premise of the Mori-Zwanzig equilibrium theory~\cite{Zwanzig1964, Zwanzig1965, Mori1965}. As we show in our simulations and experiments, this agreement stems from extracting friction from non-equilibrium drag on the particle in simulations or the sedimentation experiments, which is in essence probing the zero frequency linear response to a net force. Notably, such extracted friction is through the Eisnstein-Sutherland equation consistent with the equilibrium diffusion coefficient of C\textsubscript{60}. The very basic  nature of this finding suggest that it is applicable very generally, and therefore could be systematically used to determine friction on the molecular scale.  Recovering the \gls{ses} equation is, however, a little more delicate – for C\textsubscript{60} it requires a small partial slip on the particle surface, as suggested by a small but not negligible Knudsen number~\cite{Cunningham1910, Li2003}.  This finding truly benefited from the choice of the solute and the solvent, as this combination allows for the independent estimate of the hydrodynamic radius. This is more challenging in systems with flexible solutes and stronger solute-solvent interactions, and is a task that will be addressed in future work.


\section*{Data Availability Statement}
The data that support the findings of this study are openly available in Zenodo at http://doi.org/10.5281/zenodo.8281244, reference number 8281244.

\begin{acknowledgments}
We acknowledge funding by the Deutsche Forschungsgemeinschaft (DFG, German Research Foundation) through Project-ID 416229255 – SFB 1411 Design of Particulate Products (subprojects A01, C04 and D01) – and project INST 90-1123-1 FUGG.
We gratefully acknowledge the scientific support and HPC resources provided by the Erlangen National High Performance Computing Center (NHR@FAU) of the Friedrich-Alexander-Universität Erlangen-Nürnberg (FAU). The hardware is funded by the DFG.

\end{acknowledgments}



\bibliography{bibliography_no-eprint}

\end{document}



\title{Supplementary information:\\
The Stokes-Einstein-Sutherland equation at the nanoscale revisited}



\author{Andreas Baer}
\affiliation{Friedrich-Alexander-Universität Erlangen-Nürnberg, Department of Physics, PULS Group, Interdisciplinary Center for Nanostructured Films (IZNF), Cauerstr. 3, 91058 Erlangen, Germany}
\author{Simon E. Wawra}
\affiliation{Friedrich-Alexander-Universität Erlangen-Nürnberg, Institute of Particle Technology (LFG), Cauerstr. 4, 91058 Erlangen, Germany}
\affiliation{Friedrich-Alexander-Universität Erlangen-Nürnberg, Interdisciplinary Center for Functional Particle Systems (FPS), Haberstr. 9a, 91058 Erlangen, Germany}
\author{Kristina Bielmeier}
\affiliation{Friedrich-Alexander-Universität Erlangen-Nürnberg, Institute of Particle Technology (LFG), Cauerstr. 4, 91058 Erlangen, Germany}
\affiliation{Friedrich-Alexander-Universität Erlangen-Nürnberg, Interdisciplinary Center for Functional Particle Systems (FPS), Haberstr. 9a, 91058 Erlangen, Germany}%
\author{Maximilian J. Uttinger}
\affiliation{Friedrich-Alexander-Universität Erlangen-Nürnberg, Institute of Particle Technology (LFG), Cauerstr. 4, 91058 Erlangen, Germany}
\affiliation{Friedrich-Alexander-Universität Erlangen-Nürnberg, Interdisciplinary Center for Functional Particle Systems (FPS), Haberstr. 9a, 91058 Erlangen, Germany}
\author{David M. Smith}
\affiliation{Ruđer Bošković Institute, Division of Physical Chemistry, Group of Computational Life Sciences, Bijenička 54, 10000 Zagreb, Croatia}
\author{Wolfgang Peukert}
\affiliation{Friedrich-Alexander-Universität Erlangen-Nürnberg, Institute of Particle Technology (LFG), Cauerstr. 4, 91058 Erlangen, Germany}
\affiliation{Friedrich-Alexander-Universität Erlangen-Nürnberg, Interdisciplinary Center for Functional Particle Systems (FPS), Haberstr. 9a, 91058 Erlangen, Germany}
\author{Johannes Walter}
\email{johannes.walter@fau.de}
\affiliation{Friedrich-Alexander-Universität Erlangen-Nürnberg, Institute of Particle Technology (LFG), Cauerstr. 4, 91058 Erlangen, Germany}
\affiliation{Friedrich-Alexander-Universität Erlangen-Nürnberg, Interdisciplinary Center for Functional Particle Systems (FPS), Haberstr. 9a, 91058 Erlangen, Germany}
\author{Ana-Sunčana Smith}
\email{smith@physik.fau.de, asmith@irb.hr}
\affiliation{Friedrich-Alexander-Universität Erlangen-Nürnberg, Department of Physics, PULS Group, Interdisciplinary Center for Nanostructured Films (IZNF), Cauerstr. 3, 91058 Erlangen, Germany}
\affiliation{Ruđer Bošković Institute, Division of Physical Chemistry, Group of Computational Life Sciences, Bijenička 54, 10000 Zagreb, Croatia}


\date{\today}


\maketitle



\section{Theory}
\subsection{Mori-Zwanzig}\zlabel{sec:si:mori-zwanzig}
A different approach to obtain the \gls{gk} relations was developed by~\citet{Mori1965}.
He successfully applied the projection operator formalism to obtain a generalized non-Markovian Langevin equation.
Following~\cite{Mori1965, ZwanzigBook}, the rate of change of a dynamical variable $\vec A$ is given as
\begin{align}\label{eq:LE-mori}
\frac{\del}{\del t}\vec A(t) &= i\mat\Omega \vec A(t) - \int_0^t \d s \mat{K}(s) \cdot \vec A(t-s) + \vec F_\textup{S}(t) \:,\\
i\mat\Omega &:= (L \vec A, \vec A)\cdot( \vec A, \vec A)^{-1} \:,\\
\mat{K}(t) &:= -(L\vec F_\textup{S}(t), \vec A) \cdot (\vec A, \vec A)^{-1} \:,\\
\vec F_\textup{S}(t) &:= e^{t(\mat{1-P})L} (\mat{1-P})L \vec A  \:.\label{eq:LE-mori:force}
\end{align}
In this set of equations, $L$ is the Liouville operator, the inner product is taken to be the equilibrium phase space average $(\vec A, \vec B)=\avg{\vec A \vec B^*}$ (a tensorial quantity), $\mat P$ is the projection onto the subspace of dynamic variables, spanned by all basis vectors contributing to $\vec A$ (the so-called relevant subspace). Finally, $\mat{1 - P}$ is its complement, i.e. the projection onto the irrelevant subspace.

Using the anti-Hermitian property of $L$ with respect to the inner product, one can write a generalized \gls{fdt}
\begin{equation}
\avg{\vec F_\textup{S}(t) \vec F_\textup{S}^*(t')} = \mat K(t - t') \cdot \avg{\vec A \vec A^*}\:,
\end{equation}
which is valid even without further assumptions on the average properties of $\vec F_\textup{S}(t)$.
With this formulation, the remaining problem is the exponential of the projection of the Liouville operator $\exp(t(\mat{1-P})L)$ in the force term \zeqref{self:eq:LE-mori:force}.

Further simplifications arise by assuming a slow process, i.e. the rate of change is of the order of a small parameter $\lambda$:
\begin{equation}
\frac{\del}{\del t} \vec A(t) = L \vec A(t) = \ord{\lambda}\:.
\end{equation}
As $i\mat\Omega$ is of order $\lambda$ and the memory kernel $\mat K$ is of order $\lambda^2$, we can now approximate the convolution with its Markovian form
\begin{equation}
\int_0^t \d s \mat{K}(s) \cdot \vec A(t-s) = \int_0^\infty \d s \mat{K}(s) \cdot \vec A(t) + \Ord{\lambda^3}\:.
\end{equation}
Using the identity $\exp((\mat{1-P})Lt) = \exp(\mat{1}Lt) + \ord\lambda$, one obtains  the memory kernel as
\begin{equation}
\mat{K}(t) = \left( e^{\mat{1}Lt}(\mat{1-P})L \vec A, (\mat{1-P})L \vec A \right) \cdot (\vec A, \vec A)^{-1} + \Ord{\lambda^3}\:,
\end{equation}
which only includes the unprojected time correlation function of the quantity $(\mat{1-P})L \vec A = L \vec A - i\mat\Omega \vec A$.
Considering the expansion up to the second order in the slow variable $\lambda$, one obtains a Langevin equation
\begin{subequations}
\label{eq:LE-approx}
    \begin{align}
        \label{eq:LE-approx-a}
        \frac{\del}{\del t}\vec A (t) &= i\mat\Omega \vec A(t) - \mat K_\infty \vec A(t) + \vec F_\textup{S}(t)\:,\\
        \mat K_\infty &= \int_0^\infty \d s \mat{K}(s)\:.
    \end{align}
\end{subequations}

If quantity $\vec A$ is the momentum $\vec p$ of a specific particle, then $\vec F_\textup{S}$ is the force resulting from all degrees of freedom of the system except the momentum of the particle itself.
In the absence of external potentials, the term $i\mat\Omega \vec p$ is zero and thus $\mat P L \vec p = i\mat\Omega \vec p = 0$.
In the limiting case of a slow process, the projected force $\vec F_\textup{S}(t)$ can be approximated by the total force $F_\textup{T}(t) = \frac{\del }{\del t}\vec p(t)$ as
\begin{equation}
    \vec F_\textup{S}(t) = e^{Lt}L\vec A + \Ord{\lambda^2} \approx L\vec A(t) = F_\textup{T}(t) \:.
\end{equation}
The memory term then yields the \gls{gk} friction coefficient
\begin{equation}
    \xi = m \mat K_\infty
\end{equation}
In principle, $\xi$ is still a tensorial quantity.
However, we expect it to be diagonal as the momentum in one direction does not produce friction in an orthogonal direction.
In an isotropic system it furthermore reduces to a scalar quantity, as all diagonal elements are equal.
We can then express $\xi$ as follows
\begin{subequations}
\label{eq:friction:facf}
    \begin{align}
        \xi &= \frac{m}{\avg{\vec p \cdot \vec p}} \int_0^\infty \d s \avg{F_\textup{S}(s) F_\textup{S}(0)} + \Ord{\lambda^3}\\
            &= \frac{1}{3\kbt} \int_0^\infty \d s \avg{F_\textup{S}(s) F_\textup{S}(0)} + \Ord{\lambda^3}\\
            &= \frac{1}{3\kbt} \int_0^\infty \d s \avg{F_\textup{T}(s) F_\textup{T}(0)} + \Ord{\lambda^3}\:,
    \end{align}
\end{subequations}
where in the second line we used the equipartition theorem $\avg{\vec p \cdot \vec p} = 3m\kbt$.
Neglecting all terms of third order and higher in $\lambda$, as is done in the original Mori-Zwanzig formalism to recover the \gls{gk} relation for the friction coefficient (third line), actually resembles the limit of a particle changing its momentum infinitely slowly.
While often referred to as the limit of a frozen particle \cite{ZoranThesis, Kowalik2019}, motion with constant momentum is also covered in this limit.

Given that it is not trivial to check for the correctness of the approximated Langevin equation~\zeqref{self:eq:LE-approx} directly, we reformulate the problem~\cite{ZwanzigBook}.
We start from equation~\zeqref{self:eq:LE-mori} for the particle momentum $\vec{p}$, where the first term on the right hand side drops out.
We multiply this equation with the particle momentum $\vec{p}$ and average over the ensemble to obtain a Volterra equation of first kind.
Using the isotropy of the system and the diagonality of $\mat K_\infty$, this can be reduced to the 1D case, such that we obtain a simpler relation, which is easily accessible from \gls{md} simulations:
\begin{equation}
\frac{\del }{\del t}\avg{p(t) p(0)}
= - \int_0^t \d s K(s) \cdot \avg{p(t-s) p(0)}\:.
\end{equation}
This results directly in equation~\zeqref{eq:LE-autocorr}.

\subsection{AUC formulas}\zlabel{sec:si:auc-formulas}
For the evaluation and interpretation of \gls{sv-auc} experiments (see Appendix~\zref{sec:si:auc-interpretation} for details), the measured concentration profiles are fitted with numerical solutions of the Lamm equation to calculate $s$ and $D$ as functions of time $t$, with the parameters $\omega$ and $r$, being the angular rotor velocity and radial distance, respectively:
\begin{equation}\zlabel{eq:Lamm}
\frac{\del c}{\del t} = D\left[\frac{\del^2 c}{\del r^2} + \frac{1}{r}\frac{\del c}{\del r}\right]-s\omega^2\left[r\frac{\del c}{\del r}+2c\right]\:.
\end{equation}
The sedimentation coefficient $s$ is defined as the sedimentation velocity normalized to the applied centrifugal acceleration.
The measured values of $s$ and $D$ are both dependent on the friction coefficient and can be used to determine the molecular weight $M$ via the Svedberg equation:
\begin{equation}\zlabel{eq:Svedberg}
M = \frac{s \kbt N_A}{D(1-\bar{\nu} \rho_\textup{S})}.
\end{equation}
Hereby, $N_A$ is the Avogadro constant, $\rho_\textup{S}$ the solvent density and $\bar{\nu}$ the \gls{psv}.


\section{Methods}
\subsection{Molecular dynamics simulations}\zlabel{sec:si:methods}
\begin{figure}[ht]
    \includegraphics{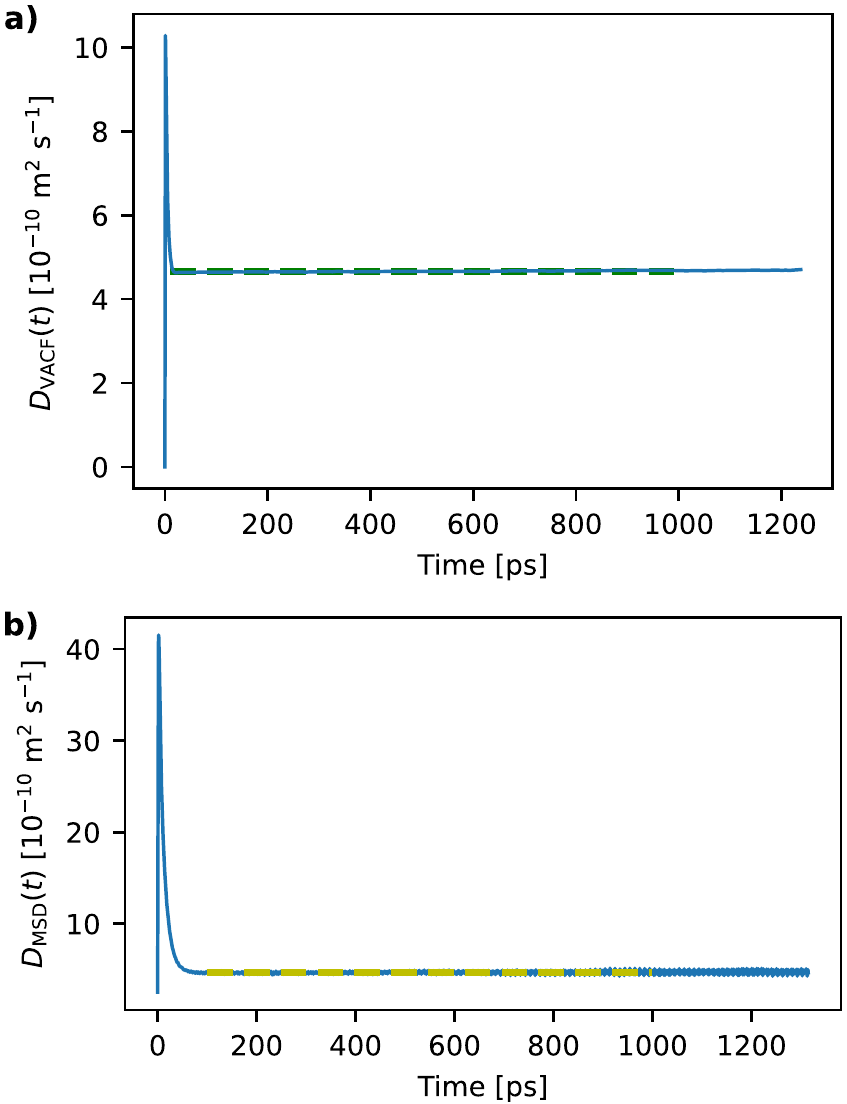}
	\caption{\textbf{Fits to obtain the diffusion coefficient.}
	The time delayed function ($D(t)$) is calculated from
	\textbf{a)} the integral of the \gls{vacf} or \textbf{b)} the time derivative of the \gls{msd}.
	Both panels show the system with a side length of \SI{8.7}{\nm} (\num{3824} toluene molecules), $\approx \SI{35}{\mu\s}$ total 3-dimensional trajectory,  before correcting for finite size effects.}
	\zlabel{fig:si:diff}
\end{figure}
We performed \gls{md} simulations of a C\textsubscript{60} dispersed in toluene (cf. Figure~\zref{fig:system}a) using the GROMACS simulation software~\cite{Gromacs1, Gromacs2, Gromacs3, Gromacs4, Gromacs5, Gromacs6, Gromacs7}, versions 2018, 2019 and 2021.
All simulations were run with a time step of \SI{1}{\fs} constraining all bonds that include hydrogen atoms using the LINCS~\cite{LINCS} algorithm. The short range non-bonded interactions were calculated using a \gls{lj} potential  switched smoothly to zero between \SIlist{0.9; 1.2}{\nm}~\cite{Christen2005}.
Long range electrostatic interactions were calculated with the \glsentryshort{pme} technique~\cite{PME1,PME2}.
For toluene, the \gls{opls}~\cite{OPLSAA1} force field is used.
For the C\textsubscript{60}, the atoms of the aromatic carbon of the \gls{opls}
force field are used together with a structure from \glsentryshort{nmr} data~\cite{Yannoni1991}.
The topology is based on the work of Monticelli~\cite{Monticelli2012}.

After minimizing the energy of the created system, a simulated annealing step is conducted at constant volume.
Hereby, initial velocities are generated according to a Maxwell distribution of the desired temperature of \SI{293.15}{\K}.
A stochastic integrator with a temperature coupling time of \SI{0.1}{\ps} is used to then heat the system up to \SI{700}{\K} over \SI{3}{\ns}, where it is kept for another \SI{3}{\ns}.
Afterwards it is cooled to the final \SI{293.15}{\K} over \SI{4}{\ns}, where it is kept for \SI{2}{\ns}.

Subsequently, an NPT equilibration run is conducted for \SI{4}{\ns}.
Hereby, a leap-frog integrator is used and temperature and pressure are controlled with a \gls{NH} thermostat and \gls{PR} barostat with coupling times of \SIlist{3.5; 8.0}{\ns}, respectively.
Afterwards, no separate NVT equilibration is run, but the first \SI{0.5}{\ns} of the production run are discarded to account for an equilibration under production conditions.

The production simulations are conducted with a leap-frog integrator with a time step of \SI{1}{\fs} and removing the \gls{com} motion every \num{10} steps.
The temperature was kept at \SI{293.15}{\K} with a \gls{NH} thermostat~\cite{NoseHoover1, NoseHoover2} with coupling every \num{10} steps and a coupling time of \SI{3.5}{\ps}.
In the steered \gls{md} simulations to calculate the friction coefficient, a constant force (with respect to the absolute coordinate system of the box) is applied to the C\textsubscript{60} to pull it through the surrounding fluid.
The relative velocity of the C\textsubscript{60} is then calculated by subtracting the instantaneous average fluid velocity from the instantaneous C\textsubscript{60} velocity.
This relative velocity is then averaged over the full simulation time and all replications.

All systems contain one C\textsubscript{60} and $N$ toluene molecules.
The \gls{gk} relations are evaluated for systems between $N=478$ and $N=46838$, corresponding to box side lengths between \SIlist{4.4; 20.2}{\nm}.
The radial distribution function was calculated for a system with \num{14907} toluene molecules, corresponding to a box size of \SI{13.8}{\nm}.

For post-processing the trajectory using \gls{gk} relations, velocities, forces and the pressure tensor are written to disk every \num{10} steps, while the positions for calculating \gls{msd} and \gls{rdf} and the velocities in the steered \gls{md} simulations are only written every \SI{200}{\fs}.
The total length of each simulation is \SI{50}{\ns} (equilibrium simulations) and \SI{10}{\ns} (steered simulations) and for each system size, between 12 and 700 realizations for each type of output are conducted.
The only exception is the simulation for the radial distribution function with three realizations of about \SI{10}{\ns} each.
Hereby, the $\approx 15000$ toluene molecules provided enough statistics to render all results statistically significant.

\subsection{Details on AUC and auxiliary measurements}\zlabel{sec:si:auc-experiments}
\subsubsection{Fullerenes - C\textsubscript{60}}
The fullerene C\textsubscript{60} with a purity of \SI{99.9}{\percent} as well as deuterated toluene at \SI{99}{atom \percent} D was purchased from Sigma Aldrich (Taufkirchen, Germany). Regular toluene with a purity of \SI{99.5}{\percent} was bought from Carl Roth (Karlsruhe, Germany). Stock solutions were prepared by dispersing C\textsubscript{60} in toluene for several days and diluting a stock solution to the desired concentration, consecutively.

\subsubsection{Experiments with MWL-AUC}
For all multi-wavelength analytical ultracentrifugation (MWL-AUC) experiments, an Optima L-90K ultracentrifuge from Beckman Coulter (Krefeld, Germany) equipped with a multiwavelength detector\cite{Walter2014} was used together with an An-60 Ti analytical rotor from Beckman Coulter and \SI{12}{\mm} measurement cell equipment from Nanolytics Instruments (Potsdam, Germany). Sedimentation velocity (SV) experiments at \SI{60}{krpm} as well as sedimentation equilibrium (SE) experiments (\SIlist{20; 30; 40; 50}{krpm})
with toluene as solvent were performed with aluminum centerpieces from Beckman Coulter. For the deuterated samples, titanium centerpieces from Nanolytics Instruments were used. The equilibrium time was greater than \SI{75}{\hour} for the SE runs. Prior to all measurements, the cells were aligned with the cell alignment tool from Nanolytics Instruments. The temperature was set to \SI{20}{\degreeCelsius} for all experiments.

\subsubsection{Experiments with Optima AUC}
\gls{sv-auc} experiments of C\textsubscript{60} were further conducted with a commercial \gls{auc}, type Optima \gls{auc}. The samples were measured at a fixed rotor speed of \SI{60}{krpm}. The temperature was kept constant at \SI{20}{\degreeCelsius} and a fixed wavelength of \SI{435}{\nm} was chosen for data acquisition. For all \gls{sv-auc} experiments, centerpieces with an optical path length of \SI{12}{\mm} were used. When converting intensity data to absorbance data, pseudo-absorption of each sample was calculated and analyzed as described by~\citet{Kar2000}.

\subsubsection{Sedimentation data evaluation}\zlabel{sec:si:auc_evaluation}
Data evaluation of \gls{sv-auc} experiments was performed with SEDFIT, version 15.01b, using the c(s)-model applying the inhomogeneous solvent model, which accounts for solvent compressibility. The finite scan velocity was also accounted for~\cite{Schuck2004}.
The analysis range was set from \SIrange{0.1}{1}{sved} (referring to water at \SI{20}{\degreeCelsius}) with at least 100 data points for discretization of the sedimentation coefficient. No regularization was applied. The frictional ratio, the meniscus position alongside time invariant (TI) and radial invariant (RI) noise were fitted by the Simplex and Marquardt-Levenberg algorithms, consecutively. The bottom position was set by optical evaluation of the first radial scan. Mean sedimentation coefficients of the main peak in the c(s)-distributions were transferred back to solvent conditions of toluene. Diffusion coefficients were calculated from the mean sedimentation coefficients and the obtained values for the frictional ratio. The parameters throughout data analysis were set to $\rho_\textup{S} = \SI{866.86}{\kg\per\m\cubed}$, $\eta = \SI{0.585}{\milli\Pa\s}$~\cite{Santos2006}, $\bar{\nu}_0 =$ \SI{4.25e-4}{\m\cubed\per\kg} and $\kappa =$ \SI{8.94e-4}{\per\MPa}.

SE experiments were evaluated with SEDANAL, version 6.93. Experiments at \SIlist{20; 30; 40; 50}{krpm} in toluene and deuterated toluene were evaluated in order to extract the buoyant mass as well as the second virial coefficient. Depending on the dataset and sample, SE-MWL data was used at wavelengths from \SI{350}{\nm} to \SI{400}{\nm}, while assuring that the extinction at the bottom of the measurement cell did not exceed a value of one. The molecular weight, the second virial coefficient, the particles’ mass concentrations and y-offsets were fitted simultaneously. In order to determine the statistical significance of the molecular weight and the second virial coefficient, F-statistics was applied with a preset confidence interval of \SI{95}{\percent}. The extinction coefficient file used for analysis was generated by the extinction coefficient at \SI{570}{\nm} and two extinction measurements on samples~\cite{Gunkin2006}.
The compressibility of the solvent could not be taken into account via the regular software options. The data was fitted with the Marquardt-Levenberg algorithm. Additionally, SV experiments were evaluated jointly with SEDANAL. In a first step, particle concentrations, $s$, $M$ as well as hydro- and thermodynamic non-ideality parameters were adjusted according to~\citet{Uttinger2019} with enabled developer options to accurately account for hydro- and thermodynamic non-ideality. After setting the concentrations as constant, F-statistics was performed for $s$ and $M$. In order to accelerate the adjustment, the radial position of the meniscus and the bottom were taken from the SEDFIT data evaluation. Using this procedure, we obtain $s_0=\SI{1.30 \pm 0.01}{sved}$ and $D_0=\SI{7.5e-10}{\m\squared\per\s}$. 

\subsubsection{Density measurements}
Density measurements of dispersions and of pure solvents were performed with a DMA 5000 M (Anton Paar, Graz, Austria) at \SI{20}{\degreeCelsius}.

\subsubsection{UV/Vis spectroscopy}
The extinctions of the particle dispersion and the particles’ mass concentrations were determined using a UV/Vis spectrometer Specord 210 (Jena Analytical, Jena, Germany). 

\subsection{Important considerations for AUC data interpretation}\zlabel{sec:si:auc-interpretation}
\subsubsection{Influence of solvent compressibility}
When analyzing \gls{auc} data acquired at high rotor speeds, a critical aspect to consider is the pressure dependency of the solvent’s viscosity and density. The effect of increasing pressure at higher radial positions and thus changing density is taken into account in the SEDFIT analysis tool using the inhomogeneous solvent model~\cite{Schuck2004} for SV data. However, the viscosity of toluene also increases for higher radial position in the measurement cell, which cannot be taken into account in SEDFIT. Therefore, one can assume that the measured sedimentation and diffusion coefficients are smaller by a factor $\epsilon = \frac{\eta_\textup{average}}{\eta_\textup{atm}}$, which is the ratio of radially averaged viscosity $\eta_\textup{average}$ and the viscosity $\eta_\textup{atm}$ at atmospheric pressure. Introducing the expressions $s_\textup{averaged}=s_\textup{atm}/\epsilon$ and $D_\textup{averaged}=D_\textup{atm}/\epsilon$ to equation~\zeqref{eq:frictional_ratio} gives a relation for the hypothetical frictional ratio $\xi/\xi_\textup{0,atm}$ at atmospheric pressure and the experimental frictional ratio $\xi/\xi_\textup{0,average}$:
\begin{equation}\zlabel{eq:si:ff0_average}
    \frac{\xi}{\xi_\textup{0,atm}}=\frac{1}{\epsilon} \frac{\xi}{\xi_\textup{0,average}}
\end{equation}
Equation~\zeqref{eq:si:ff0_average} indicates that an increase in viscosity influences both sedimentation as well as diffusion and thus leads to smaller values for the frictional ratio. During an \gls{auc} experiment, a hydrostatic pressure gradient develops over the entire cell. The pressure difference strongly depends on the compressibility of the solvent~\cite{Schuck2004, Uttinger2019}.
In the case of toluene with a solvent compressibility of \SI{8.94e-10}{\per\Pa}, the relative nonlinear change in the solvent viscosity amounts up to \SI{19}{\percent}~\cite{Uttinger2019}. Therefore, the interpolation to small rotor speeds, hence atmospheric pressure, is slightly affected. A typical value for toluene interpolated to \SI{20}{\degreeCelsius} would be $\epsilon = 1.03$ based on the calculation of the pressure within an \gls{auc} cell filled with a compressible solvent~\cite{Schuck2004, Harris2000}.
Therefore the value of the frictional ratio, which is obtained from extrapolation to atmospheric pressure, is expected to be \SI{3}{\percent} lower than given directly by $s$, $D$ and equation~\zeqref{eq:frictional_ratio}. 

\subsubsection{Influence of bottom position on sedimentation coefficients}
Besides limited detection capabilities for large concentration gradients, the analysis of measured sedimentation boundaries can be prone to error due to the bottom position affecting the outcome of the data analysis. This is especially important when dealing with small particles and narrow density contrasts as the concentration distribution in the SE covers the entire sedimentation path. The radial SE concentration distribution is the limit of each \gls{sv-auc} experiment. Therefore, if the radial concentrations of the SE profile are non-zero in the analysis interval, the correct setting of the bottom position is imperative as it influences data evaluation. This can be explained by the fact that back-diffusion depends on the exact position of the bottom. SEDFIT’s c(s)-model allows setting or fitting the bottom position. However, for the experimental data investigated here, fitting leads to deviations from the actual bottom position, which translates into minor uncertainties in the sedimentation coefficient distributions. In addition, a small peak (signal $<\SI{5}{\percent}$) is found after data analysis, as can be seen in Figure~\zref{fig:si:bottom}. This artifact is also present in the work of~\citet{Pearson2018} with a fit of the bottom position. The occurrence of the peak is mostly due to the small sedimentation fluxes with respect to high diffusivity of C\textsubscript{60} leading to insufficient meniscus depletion.

\begin{figure}[ht]
    \includegraphics[width=\columnwidth]{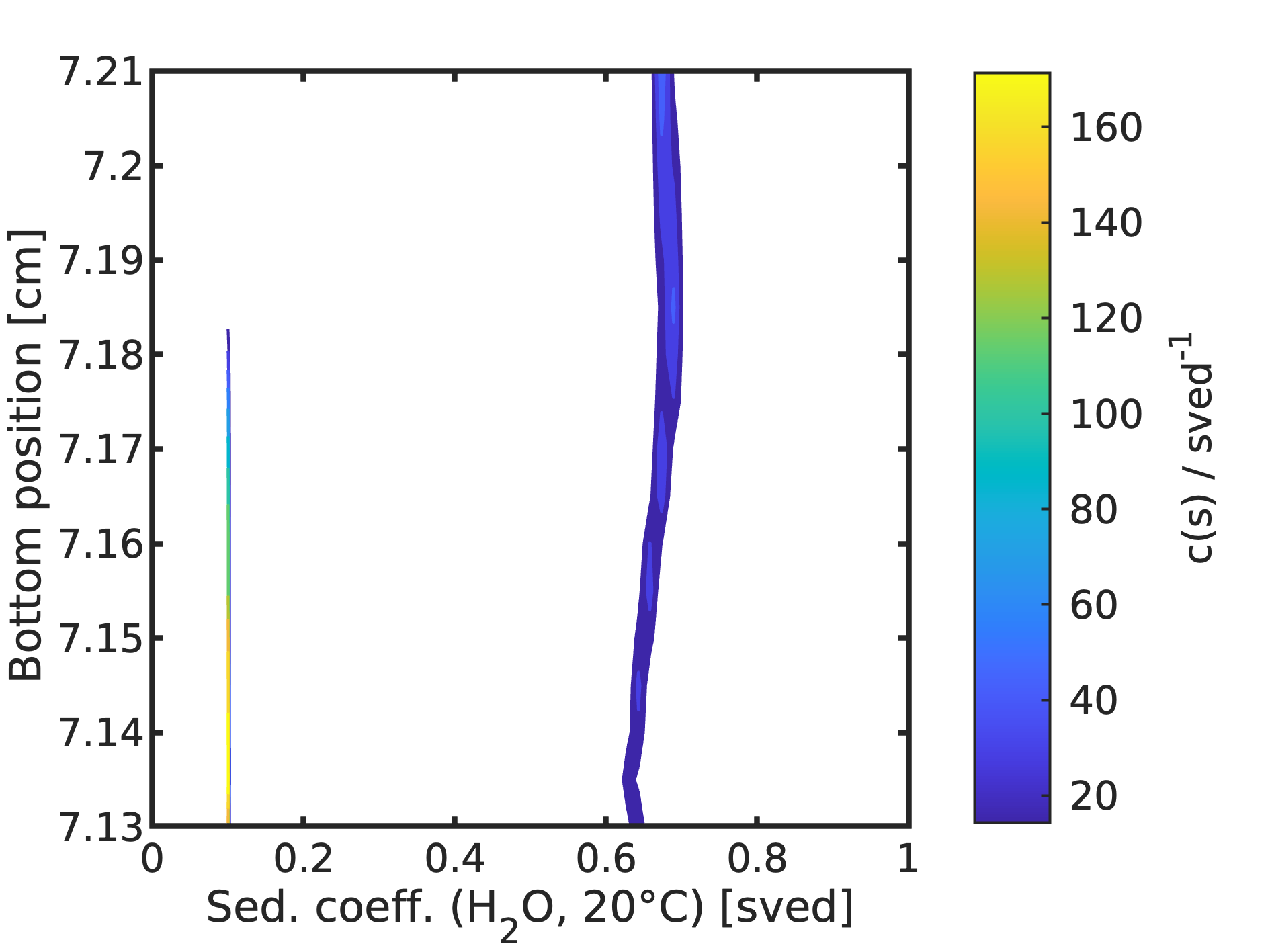}
	\caption{\textbf{Sedimentation coefficient distribution $c(s)$ as a function of bottom position $r_\textup{B}$} The values are retrieved from the inhomogeneous solvent model in SEDFIT~\cite{Schuck2004}.
	}
	\zlabel{fig:si:bottom}
\end{figure}   

Here, we vary the bottom position in order to the lowest associated root-mean-square deviation (RMSD). However, as can be seen in Figure~\zref{fig:si:bottom:fit}, the fitted value of the bottom position does not correspond to the position found in the first radial scan. Instead of fitting $r_\textup{B}$, together with $\xi/\xi_0$, $r_\textup{M}$ as well as TI and RI noise, it can be also fixed, allowing the extraction of the RMSD($r_\textup{B}$). It is obvious that the Marquardt-Levenberg algorithm finds the minimum RMSD well, while the real bottom position corresponds to a higher RMSD. In summary, our data analysis shows that a correct setting of the bottom positions limits the occurrence of artifact peaks at small sedimentation coefficients for our \gls{sv-auc} data.

\begin{figure}[ht]
    \includegraphics{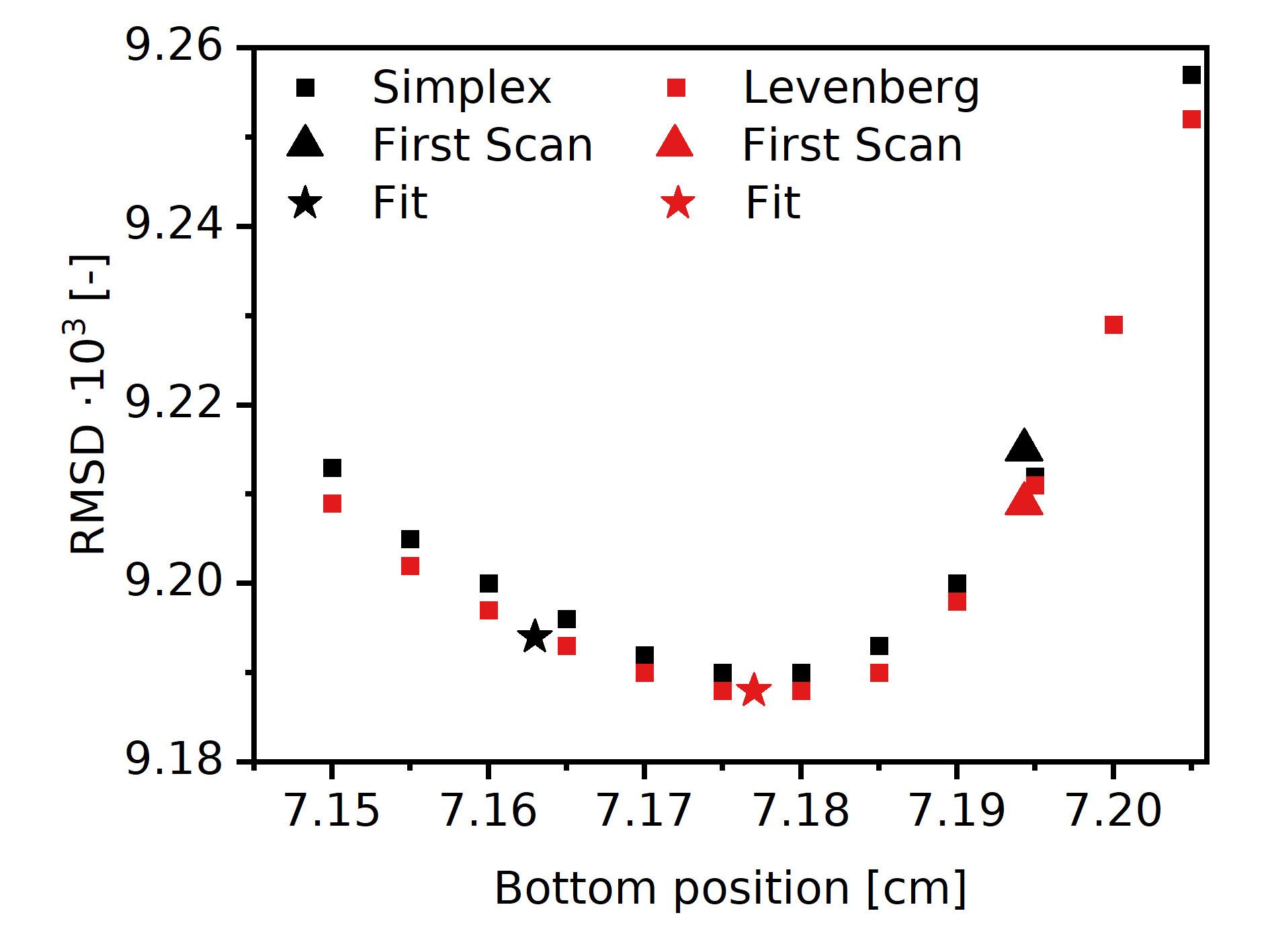}
	\caption{\textbf{RMSD values as a function of the bottom position.} Hereby, the bottom position was set and the meniscus position and $\xi/\xi_0$ are fitted, along RI and TI noise within the c(s) method.}
	\zlabel{fig:si:bottom:fit}
\end{figure}   

\subsubsection{Partial specific volume}\zlabel{sec:si:psv}
The \gls{psv} is relevant for the full hydrodynamic and thermodynamic analysis of \gls{auc} data. Experimentally, the \gls{psv} can be either determined by the Kratky method~\cite{Kratky1973} or via \gls{auc} using density contrast (DC) measurements (e.g., using deuterated and non-deuterated solvents).

\subsubsection{Determination of the PSV using the Kratky method}
For the Kratky method, the density increment of dispersions with different analyte mass concentrations is measured and the \gls{psv} is determined according to:
\begin{equation}
    \bar{\nu}=\frac{1-\del\rho/\del c}{\rho_\textup{S}},
\end{equation}
 with $\rho$ being the solution density, $\rho_\textup{S}$ the solvent density and $c$ the analyte mass concentration. Typically, the concentrations of the dispersions are determined via UV/Vis extinction measurements and conversion of the extinction $E$ to mass concentration is obtained by a known mass extinction coefficient. The concentration is then determined via the Lambert-Beer law at wavelength $\lambda$ and with the optical path length $l$:
\begin{equation}
E(\lambda) = \log\left(\frac{I_0(\lambda)}{I(\lambda)}\right)=\epsilon(\lambda) c l
\end{equation}

Here, we use literature values for the value of the extinction coefficient $\epsilon(\lambda)$. Normalizing the measured extinction coefficients to a defined wavelength permits the evaluation of the accordance between the extinction coefficients given at different wavelengths, as can be seen in Figure~\zref{fig:si:extinction_norm}. While the values from~\citet{Gunkin2006} and~\citet{Sension1991} match well, the value from~\citet{Barroso1998} shows an offset. This may be either due to experimental uncertainty or due to a systematic mismatch. It is noteworthy that all values are given as molar extinction coefficients. However, as stated in their work, the amount of C\textsubscript{60} was determined via weighting, therefore calculating back to a mass extinction coefficient seems valid and should correspond to the originally obtained dataset.

\begin{figure}[ht]
    \includegraphics{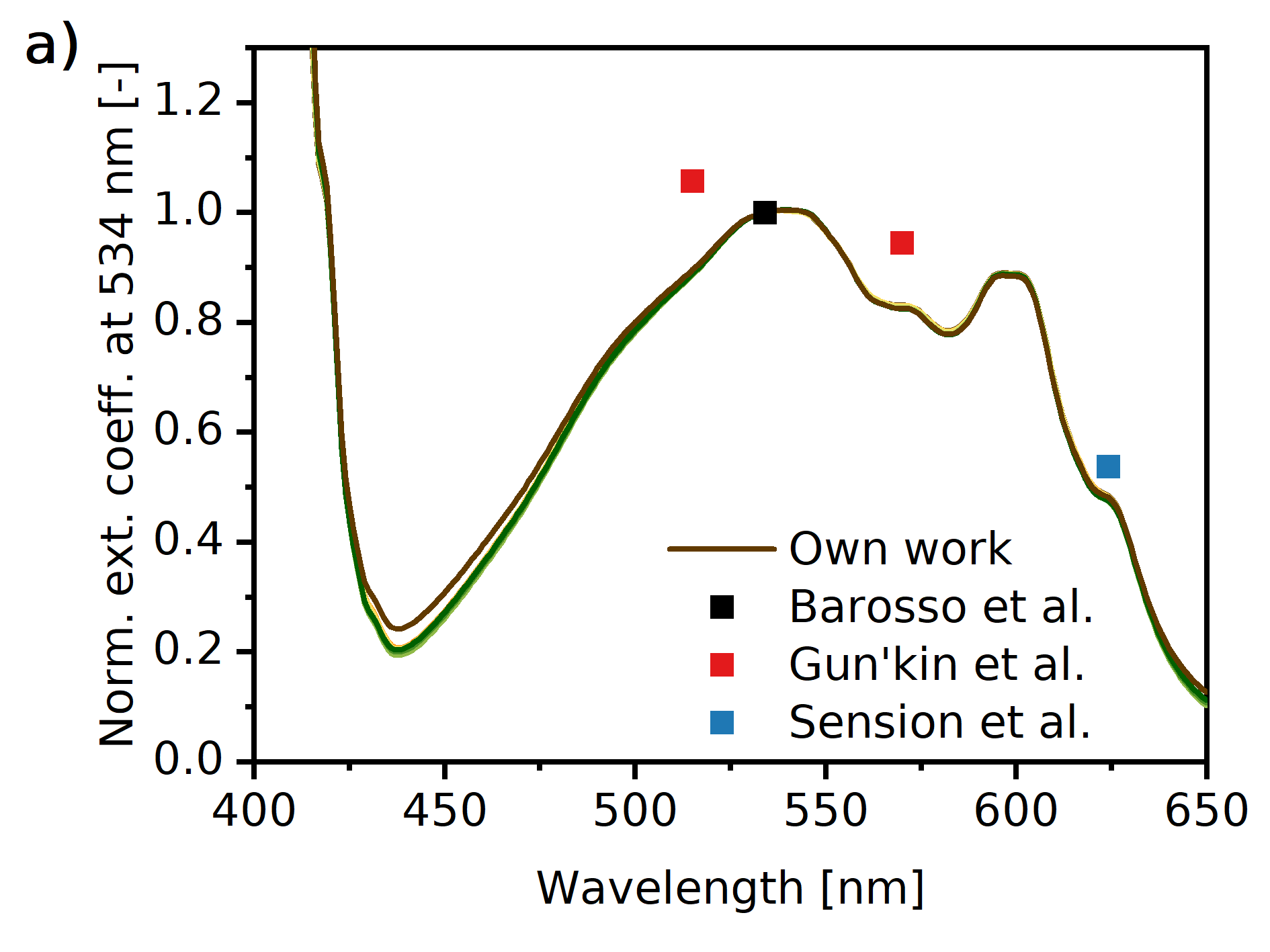}
    \includegraphics{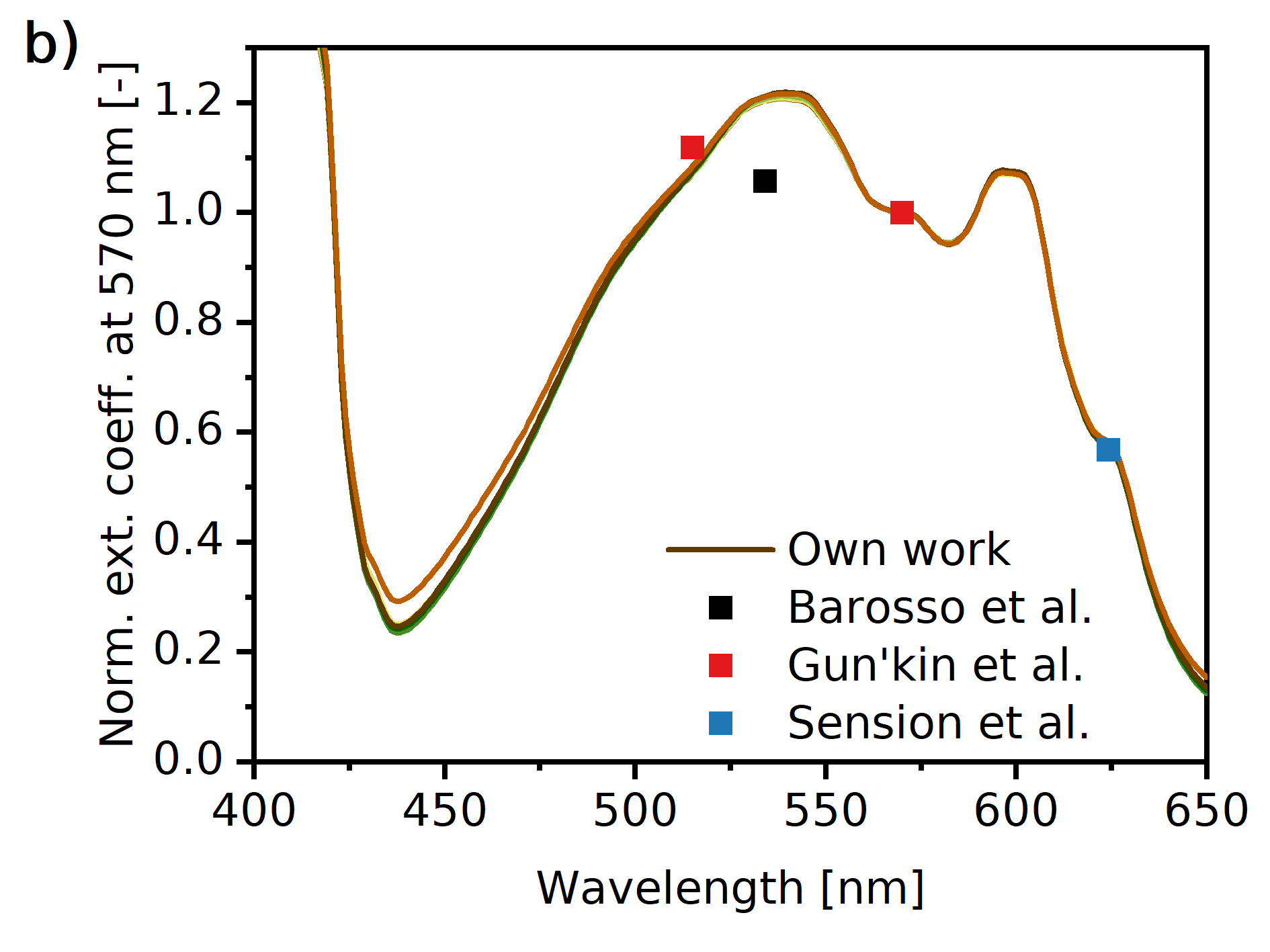}
	\caption{\textbf{Normalized extinction coefficients retrieved from literature}~\cite{Sension1991, Barroso1998, Gunkin2006}. Normalized to \textbf{a)} \SI{534}{\nm} and \textbf{b)} \SI{570}{\nm}. The original spectra are displayed in Figure~\zref{fig:si:extinction_raw}.}
	\zlabel{fig:si:extinction_norm}
\end{figure}

The \gls{psv} is then determined by calculating the concentration via the measured extinction spectra and using the extinction coefficient value at \SI{570}{nm} and \SI{534}{nm}~\cite{Sension1991, Barroso1998, Gunkin2006}.
The measured extinction spectra and the resulting densities for the two selected wavelengths are depicted in Figure~\zref{fig:si:extinction_raw} and \zref{fig:si:extinction_correlation}. It is advantageous for future studies to give the results extinction based as $\del \rho / \del E$ instead of concentrations based as $\del \rho / \del c$, thus allowing reinterpretation of data and uncertainties with extinction coefficients of higher accuracy.

\begin{figure}[ht]
    \includegraphics{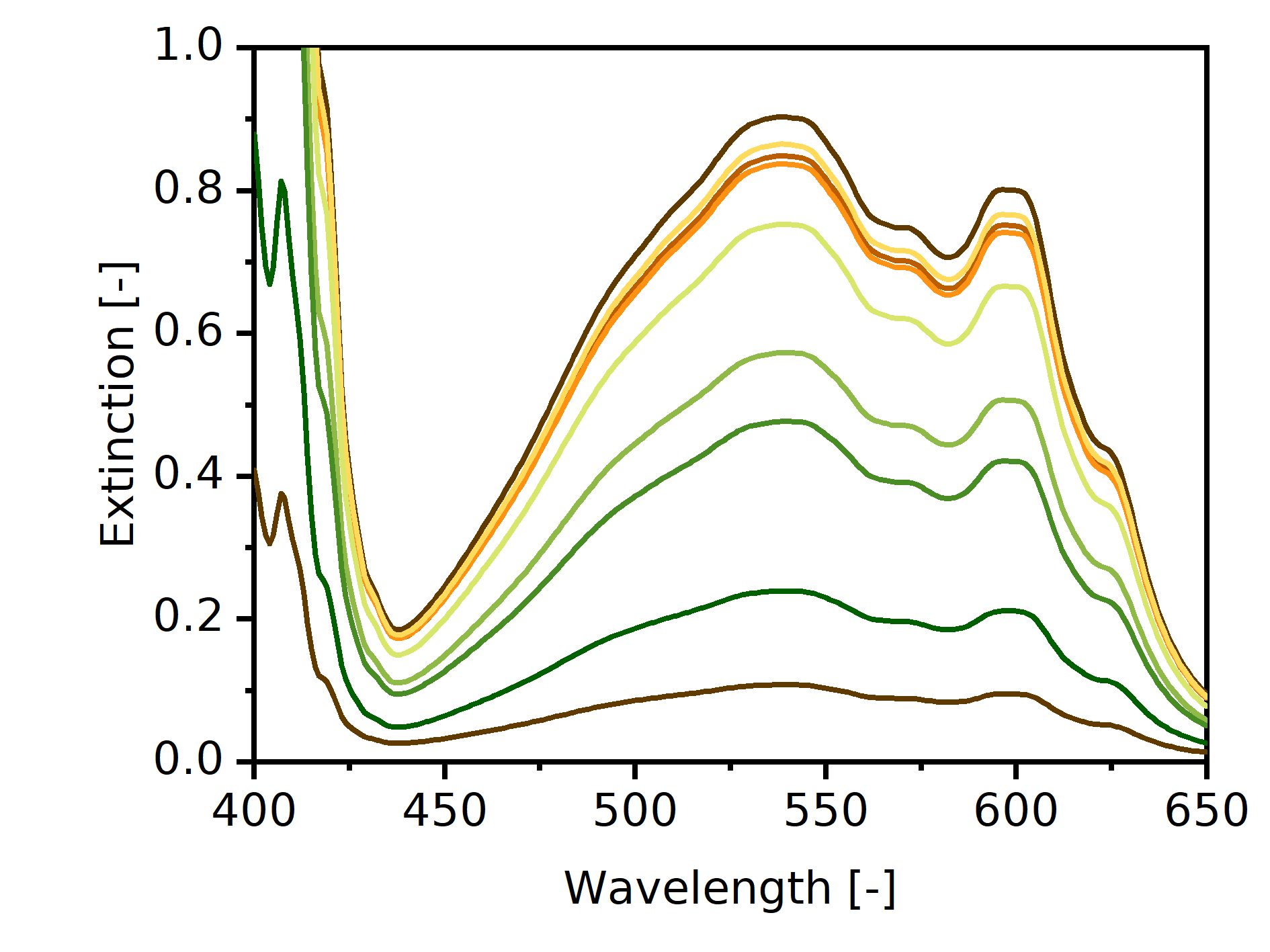}
	\caption{\textbf{Extinction spectra of dispersions used for density measurements.} Please note that extinction was calculated using the decadic logarithm and an optical path length of \SI{1}{\cm}.}
	\zlabel{fig:si:extinction_raw}
\end{figure}

\begin{figure}[ht]
    \includegraphics{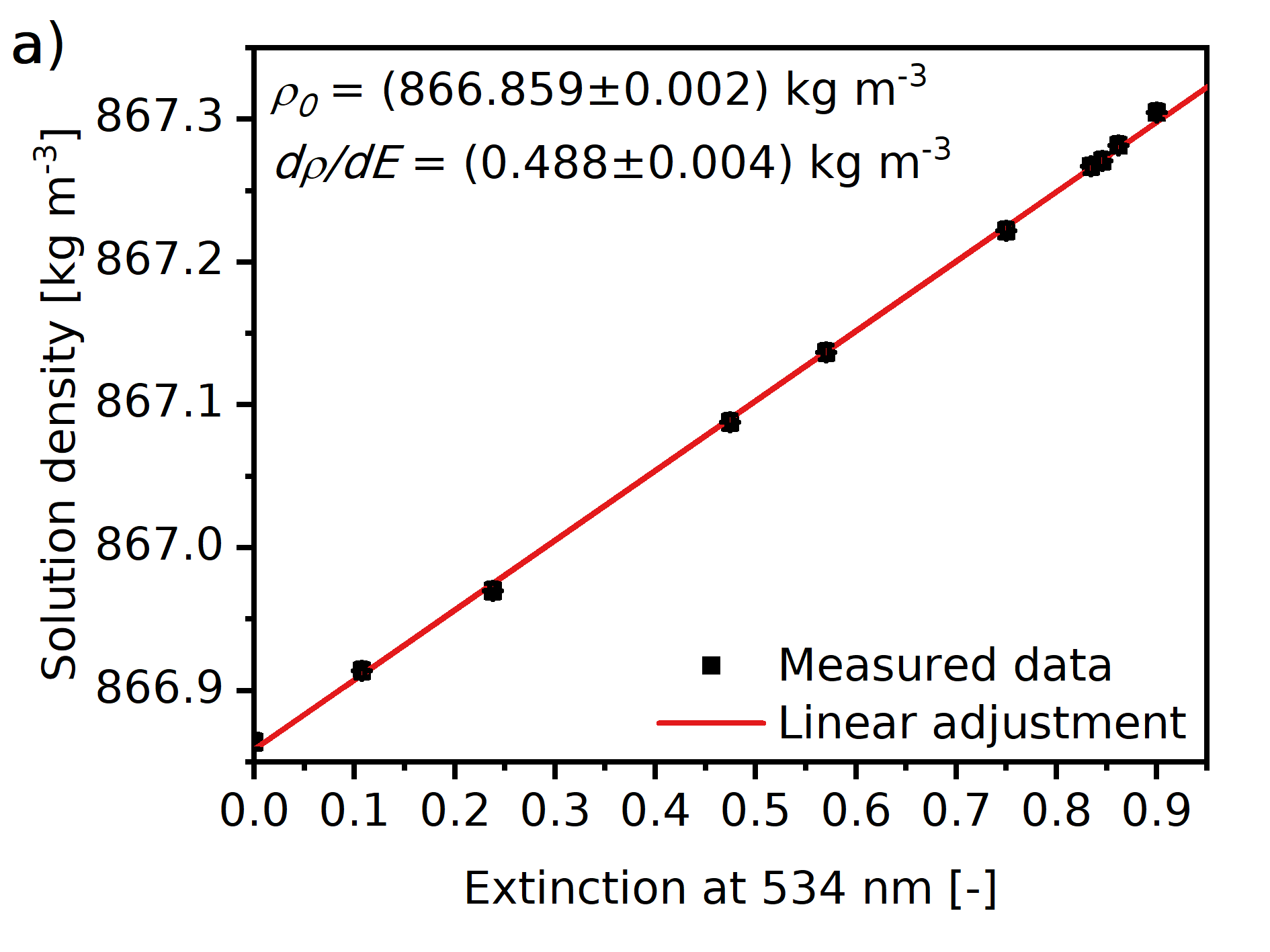}
    \includegraphics{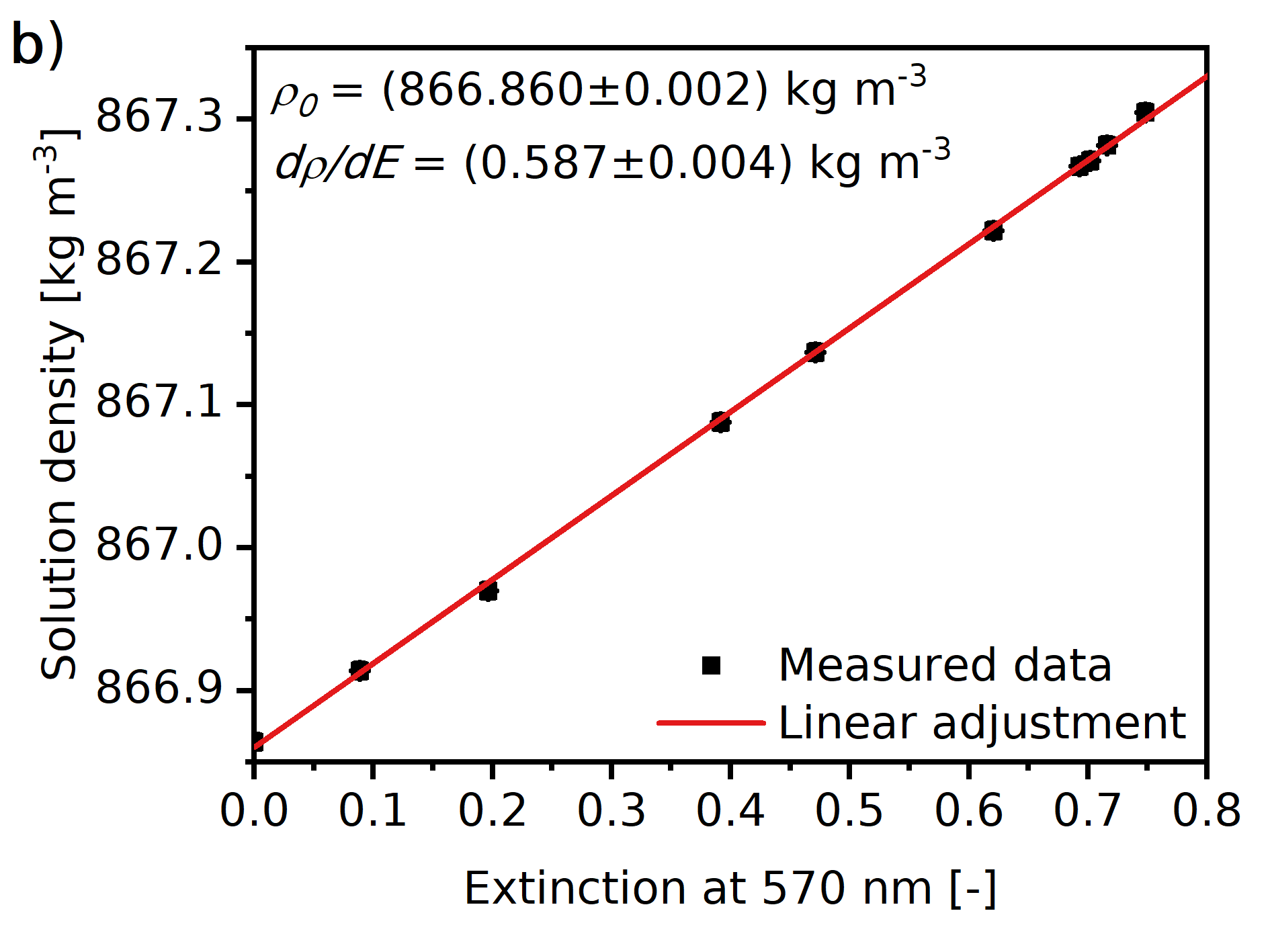}
	\caption{\textbf{Extracted extinction values and determined dispersion density.}  The values are for wavelengths of \textbf{a)} \SI{534}{\nm} and \textbf{b)} \SI{570}{\nm} and the raw data is depicted in Figure~\zref{fig:si:extinction_raw}.
	Error bars are given for density as well as extinction and are included in the fitting procedure in Gnuplot to evaluate the uncertainties of $\rho(E=0)$ and $\del\rho/\del E$.}
	\zlabel{fig:si:extinction_correlation}
\end{figure}
  
In order to calculate the uncertainty of the resulting \gls{psv}, the determining equation has to be expanded in order to provide a direct access to all relevant parameters:

\begin{equation}
   \bar{\nu}_\textup{Kratky} = \frac{1}{\rho_\textup{S}}\left(1-\epsilon l \frac{\del \rho}{\del E}\right)
\end{equation}

In order to obtain an uncertainty of $\del \rho / \del E$, the uncertainties of the individual measured densities and extinction values given by the manufacturers (\SI{0.00502}{\kg\per\m\cubed} and \num{0.003}, respectively) were taken into account.\\
The necessary derivatives for the propagation of uncertainty are as follows:

\begin{equation}
    \frac{\del \bar{\nu}_\textup{Kratky}}{\del \rho_\textup{S}} = - \frac{1}{\rho_\textup{S}^2}\left(1-\epsilon l \frac{\del \rho}{\del E}\right) = - \frac{\bar{\nu}_\textup{Kratky}}{\rho_\textup{S}}
\end{equation}

\begin{equation}
    \frac{\del \bar{\nu}_\textup{Kratky}}{\del \epsilon} = - \frac{1}{\rho_\textup{S}} l \frac{\del \rho}{\del E}
\end{equation}

\begin{equation}
    \frac{\del \bar{\nu}_\textup{Kratky}}{\del \frac{\del \rho}{\del E}} = - \frac{1}{\rho_\textup{S}} \epsilon l
\end{equation}

\begin{equation}
    \frac{\del \bar{\nu}_\textup{Kratky}}{\del l} = - \frac{1}{\rho_\textup{S}} \epsilon \frac{\del \rho}{\del E}
\end{equation}

With the additional uncertainties of $u_l =$ \SI{0.001}{\cm} and an estimated uncertainty of $u_{\epsilon(\SI{570}{\nm})} = 0.03\epsilon(\SI{570}{\nm})$ (estimated using values from Gun’kin and Sension) from Figure~\zref{fig:si:extinction_norm}, one obtains:

\begin{equation}
    \bar{\nu}_\textup{Kratky, \SI{570}{\nm}} = \SI{4.26 \pm 0.22 e-4}{\m\cubed\per\kg}
\end{equation}

For the value of the \gls{psv} at \SI{534}{\nm}, it is not possible to calculate any uncertainty from the given data, therefore the uncertainty from \SI{570}{\nm} is adapted $u_{\epsilon(\SI{534}{\nm})} = 0.03 \epsilon (\SI{534}{\nm})$, which gives:

\begin{equation}
    \bar{\nu}_\textup{Kratky, \SI{534}{\nm}} = \SI{5.14 \pm 0.22 e-4}{\m\cubed\per\kg}
\end{equation}

The corresponding volume equivalent radii are $R_\textup{Kratky, \SI{570}{\nm}} = \SI{0.495+-0.009}{\nm}$ and $R_\textup{Kratky, \SI{534}{\nm}} = \SI{0.528+-0.007}{\nm}$ as described in SI section~\zref{sec:si:radius:static}.

\subsubsection{Determination of the PSV using DC SV- and SE-AUC}
Another way of determining the \gls{psv} is the use of \gls{auc} experiments with varying amounts of deuterated solvents, thus changing both solvent density as well as solvent viscosity~\cite{Edelstein1967, Maechtle1984}.

Under the assumption of negligible preferential absorption, there is no effect of solvation on the resulting \gls{psv} from two different SV runs:
\begin{equation}
\bar{\nu}_\textup{SV}=\frac{s_1 \eta_1 - s_2 \eta_2}{s_1 \eta_1 \rho_\textup{S2}-s_2 \eta_2 \rho_\textup{S1}}
\end{equation}
While \gls{sv-auc} experiments depend on both solvent density and solvent viscosity, SE-AUC solely depends on the solvent density. This is advantageous as a reduced parameter set leads to smaller propagation of uncertainties. Additionally, by using SE, one bypasses the problem of incomplete radial depletion for the precise determination of the sedimentation coefficient. Analysis tools like SEDANAL~\cite{Sherwood2016} allow accessing the buoyant molecular weight $M_\textup{B}=M(1-\bar{\nu}\rho_\textup{S})$. The \gls{psv} can then be calculated from determined values of $M_\textup{B}$ in a non-deuterated solvent $M_\textup{B1}$ and a deuterated solvent $M_\textup{B2}$. With the boundary condition of equivalent molecular weights, it follows:
\begin{equation}
\bar{\nu}_\textup{SE}=\frac{M_\textup{B1} - M_\textup{B2}}{\rho_\textup{S2} M_\textup{B1} - \rho_\textup{S1} M_\textup{B2}}
\end{equation}
Using SEDANAL and including the results from the F-statistics calculations, the \gls{psv} was determined to be \SI{5.10+-0.40e-4}{\m\cubed\per\kg}. The corresponding volume equivalent radius is $R_V^\textup{exp, AUC} = \SI{0.526+-0.014}{\nm}$ \\
Notably, a prominent factor for the determination of the \gls{psv} from SE experiments is the solvent compressibility. Moreover, an interference of solvent compressibility and thermodynamic non-ideality through the second virial coefficient cannot be fully excluded, as these parameters directly influence the sedimentation and diffusional properties of the fullerenes~\cite{Uttinger2019}.


\section{Results}
\subsection{Retrieving the diffusion coefficient}\zlabel{sec:si:diffusion}
\subsubsection{Finite size effects}\zlabel{sec:si:finite-size}

\gls{md} simulations unavoidably violate the thermodynamic limit and finite size effects are always present.
This problem is reduced with periodic boundary conditions and long sampling times.
However, issues persists even in this case~\cite{Duenweg1993, Yeh2004, Alper2021}.
The problem of finite size was addressed using hydrodynamic theory by~\citet{Duenweg1993} as well as~\citet{Yeh2004}, who calculated the dependence of the diffusion coefficient on the system size \zeqref{eq:diff-size_dependence}, when periodic boundary conditions apply.
Here, $D_\infty \equiv D$ is the true and $D_\textup{PBC}$ the apparent diffusion coefficient, while $a$ is the characteristic system size. The constant $\zeta$ is the analogue to a Madelung constant and related to the type of lattice that the periodic replications of the simulation box produce.

For a cubic box, $a$ adopts the value of the box side length, and $\zeta = \num{2.837297}$ as found by several authors in the past.
The coefficient $b$ is the coefficient for the boundary condition, varying between the two extremes of $b=4$ and $b=6$ for slip and stick, respectively.
In the original derivation~\cite{Duenweg1993, Yeh2004}, stick boundary conditions were used throughout.
However, the coefficient is directly inherited from the Oseen tensor, that is stated with stick boundary conditions, thus a variable coefficient can be used throughout the whole calculation.

Similarly, following the derivation of~\citet{Yeh2004}, the friction coefficient $\xi_\infty \equiv \xi$ can be extracted from the apparent friction coefficient $\xi_\textup{PBC}$  evaluated in \gls{md} simulations as
\begin{equation}\zlabel{eq:friction-size}
\frac{1}{\xi_\infty} = \frac{1}{\xi_\textup{PBC}} + \frac{\zeta}{b\pi \eta a}
\end{equation}
These corrections will be systematically applied to the presented data, as appropriate, i.e. all references to $D$ and $\xi$ refer to the value of infinite system size.

\subsubsection{Reference viscosity}\zlabel{sec:si:diffusion:viscosity}
\begin{figure}
    \includegraphics{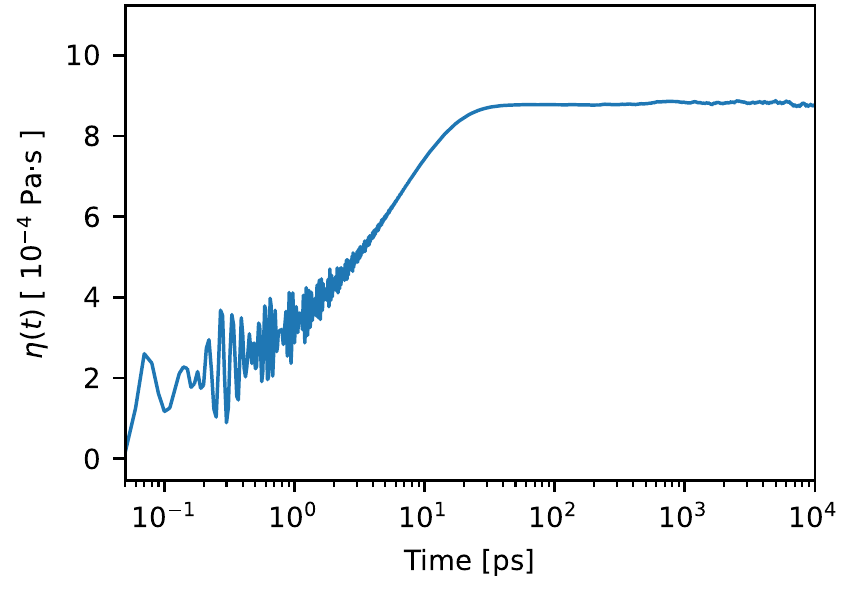}
	\caption{\textbf{Solvent viscosity in \gls{md} simulations.}
	The integral of the stress autocorrelation function is shown as a function of delay time $t$, the long time limit of which is the solvent shear viscosity $\eta$.
	The data is sampled every \SI{10}{\fs} and averaged over all six off-diagonal elements of the stress tensor.
	Furthermore, it is averaged over all simulations and all times, accounting for a total of about \SI{54}{\micro\s}.
	The plateau value is sampled with high precision for more than one order of magnitude in time to ensure proper convergence.
	The presented data is for the system with a side length of \SI{8.7}{\nm}.
	}
	\zlabel{fig:results:visco}
\end{figure}
Given that the shear viscosity explicitly enters the \gls{ses} equation, in \gls{md} simulations $\eta$ is obtained independently by utilizing its relation to the stress
autocorrelation function through the appropriate \gls{gk} relation:
\begin{equation}
\eta = \frac{V}{\kBT} \int_0^\infty \avg{P_{\alpha\beta}(t_0 + t) P_{\alpha\beta}(t_0)} \d t\:.
\end{equation}
Hereby, $V$ is the volume of the simulation box and the average runs over all off-diagonal elements of the stress tensor $P_{\alpha\beta}\;(\alpha,\beta = x,y,z\;:\; \alpha \neq \beta)$, thus all index pairs.
As expected~\cite{Yeh2004, Milicevic2014} the system size dependence is small, thus the mean over all systems is taken for the analysis, yielding $\eta = \SI{8.48 \pm 0.08 e-4}{\Pa\s}$ (cf. Figure~\zref{fig:results:visco}).
The viscosity is extracted from the same simulations from which the \gls{vacf} and the \gls{ftacf} are extracted, details for which can be found in section~\zref{sec:si:methods}.

The value for the viscosity of toluene is significantly larger than the experimentally measured $\eta = \SI{5.9 e-4}{\Pa\s}$~\cite{Santos2006, Pearson2018}.
This is an issue of the \gls{opls} force field~\cite{OPLSAA1}, that has been used to parameterize toluene.
This model, as well as most standard force fields, have been developed to accurately recover the fluid density and its structure factor among other properties, whereas the emphasis on dynamic characteristics is somewhat smaller. Consequently, the viscosity is often not perfectly recovered.
While this discrepancy should be taken into account when comparing the absolute diffusivities to experimental measurements, it will not affect our assessment of the validity of the \gls{ses} equation. 

Equations~\zeqref{eq:intro:einstein} to \zeqref{eq:intro:ses} suggest that the product of diffusivity and viscosity is constant. For simulations of toluene, this product is
\begin{align*}
    D_\textup{tol}^\textup{sim} \cdot \eta_\textup{tol}^\textup{sim}
    & = \SI{15.1 e-10}{\m\squared\per\s} \cdot \SI{8.48 e-4}{\Pa\s} \\
    & = \SI{1.28}{\pico\N}\:.
\end{align*}
Using available experimental data~\cite{Oreilly1972, Santos2006}, the same product is 
\begin{align*}
    D_\textup{tol}^\textup{exp} \cdot \eta_\textup{tol}^\textup{exp}
    & = \SI{21.2e-10}{\m\squared\per\s} \cdot \SI{5.88 e-4}{\Pa\s} \\
    & = \SI{1.25}{\pico\N}\:,
\end{align*}
showing that the simulation result has only a \SI{3}{\percent} deviation for the product.

Using this fact, we compare the experimentally available data for diffusivities of C\textsubscript{60} with simulations by calculating a rescaled diffusion coefficient $D_\textup{C\textsubscript{60}}^\textup{sim}\cdot\eta_\textup{tol}^\textup{sim}/ \eta_\textup{tol}^\textup{exp}$ (cf. column 4 in Table~\zref{tab:results:summary}).

\subsubsection{Diffusivity from positional and velocity correlations in MD simulations}\zlabel{sec:si:diffusion:vacf}
We start by calculating the diffusion coefficient of C\textsubscript{60} from its \gls{vacf} using the \gls{gk} relation, as established by~\citet{Alder1970} (cf. Figure~\zref{fig:results:diffusion}a):
\begin{equation}
D_\textup{\glsentryshort{vacf}} = \int_0^\infty \avg{v(t_0 + t) v(t_0)} \d t\:.
\end{equation}

To ensure proper sampling and evaluation of finite size effects, 6 systems were studied with \num{478} to \num{46838} solvent molecules and sizes $a = \SIrange{4.4}{20.2}{\nm}$.
At least 12, but up to 700 realizations of \SI{50}{\ns} trajectories of C\textsubscript{60} were created in simulations for each system size, with the velocity of C\textsubscript{60} extracted every \SI{10}{\fs}. This provided in total \SIrange{0.6}{35}{\micro\s} of trajectory, being additionally averaged over all three dimensions.
In order to speed up data processing, the output routine of GROMACS was slightly modified to restrict the trajectory output to the particle, instead of writing the velocities for the entire system including the solvent molecules.

The \gls{vacf} was calculated for each trajectory individually (averaging over all $t_0$) and then averaged over all realizations and the three spatial dimensions before integration.
The diffusion coefficient is obtained as a fit through the plateau of the obtained running integral from \SIrange{100}{1000}{\ps} (cf. SI-Figure~\zref{fig:si:diff}a).
The error is calculated as the root mean squared deviation of the plateau data from the fit.

As most commonly performed in \gls{md} simulations, the diffusion coefficient can be derived from the time derivative of the \gls{msd} (Figure~\zref{fig:results:diffusion}b) via
\begin{equation}
D_\textup{\glsentryshort{msd}} = \lim_{t \rightarrow \infty} \frac{1}{2}\frac{\d}{\d t}\avg{[x(t_0 + t) - x(t_0)]^2}\:.
\end{equation}
The \gls{msd} is obtained from the same simulations as the \gls{vacf} with the positions saved every \SI{200}{\fs} as suggested by~\citet{Milicevic2014}.
The diffusion coefficient is obtained in an analogue way by fitting the plateau value of the time derivative of the \gls{msd} (cf. SI-Figure~\zref{fig:si:diff}b).

From a theoretical point of view, both $D_\textup{\glsentryshort{vacf}}$ and $D_\textup{\glsentryshort{msd}}$ should be identical, however differences in convergence should and have been observed. This was first clarified by~\citet{Alder1970} who also showed that the velocity autocorrelation function decays as $t^{-3/2}$ in the long time limit due to hydrodynamic effects. An indication for such a decay is now found in our molecular system (Figure~\zref{fig:results:diffusion}a), albeit a large sampling and significant system size is required to make this observation. In terms of extracted diffusivities, we also obtain coinciding results with high accuracy from both techniques (Figure~\zref{fig:results:diffusion}c). Nonetheless, the statistical uncertainties are smaller for the \gls{msd}.

\subsubsection{D from FACF}\zlabel{sec:si:diffusion:facf}
\begin{figure}
    \includegraphics{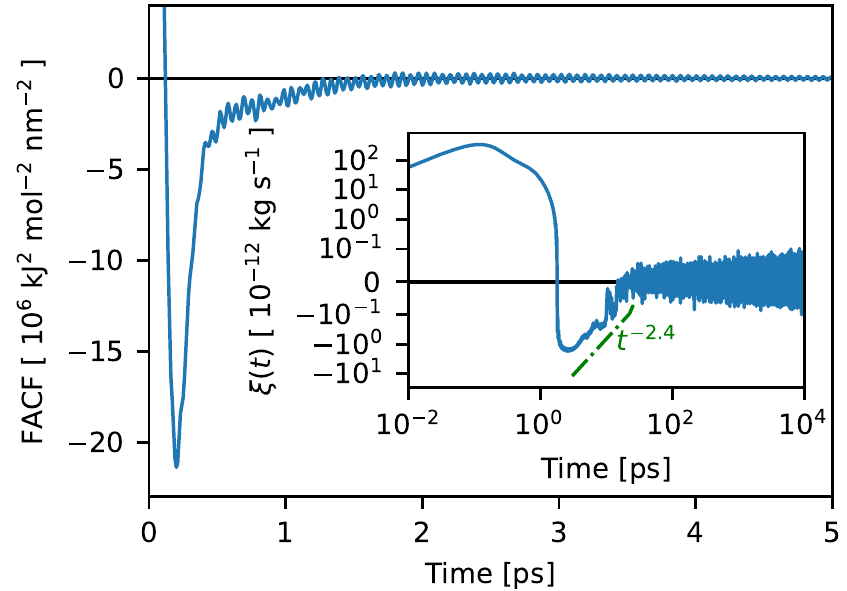}
	\caption{\textbf{Friction coefficient obtained from the \glsentrylong{facf}.}
	The \gls{ftacf} for C\textsubscript{60} is shown as a function of delay time $t$.
	The inset shows the running integral of the \gls{ftacf} following Kirkwood \zeqref{self:eq:friction:facf}. The experimental friction coefficient is to be calculated as the plateau value, or from the exponential decay, but neither is present (see text for details).
	Instead, a power law decay consistent with the expected value of $t^{-5/2}$ is observed.}
	\zlabel{fig:results:facf}
\end{figure}
An alternative approach to calculate the diffusion constant is using the Einstein-Sutherland equation~\zeqref{eq:intro:einstein}, whereby the friction coefficient is calculated from the \gls{ftacf}~\zeqref{self:eq:friction:facf}.
Practically, the friction coefficient is calculated from the same simulations as the diffusion coefficient in section~\zref{sec:si:diffusion:vacf} with \SIrange{0.6}{35}{\micro\s} total simulation time and output rate of the total force acting on the C\textsubscript{60} of \SI{10}{\fs}. The latter step required the same adjustment of the output routine of GROMACS. As we have an isotropic system, we can perform the averaging of the \gls{ftacf} over all spatial dimensions. This also eliminates the normalization $1/3$ for the three spatial dimension in equation~\zeqref{self:eq:friction:facf}.

The problem with this approach, as pointed out already by~\citet{Kirkwood1946}, is that the running integral of the \gls{ftacf} approaches zero as the integration time approaches infinity due to time reversibility of hydrodynamics at these scales.
Kirkwood proposed that for all relevant time scales of observation there shall be a plateau value of the running integral at which the integration can be truncated to yield the correct friction coefficient, as measured experimentally. The first \gls{md} simulations on this topic, however, found no such plateau value~\cite{Lagarkov1978}. Consequently, the authors used the maximum in the integral occurring at the first zero-crossing of the \gls{ftacf} as an approximation of the friction coefficient.

Later simulations proposed that the decay shall be exponential and they successfully obtained the friction coefficient from the decay rate~\cite{Espanol1993, Ould-Kaddour2003}.
A systematic analysis of the influence of the mass of a soft Lennard-Jones particle suspended in SPC/E water on the \gls{ftacf}~\cite{ZoranThesis} showed, that for particle to solvent mass ratios up to \num{100}, the \gls{ftacf} is dominated by backscattering effects.
However, an exponential decay of the running integral of the \gls{ftacf} proposed by~\citet{Espanol1993, Ould-Kaddour2003} was found for heavier particles.
Interestingly, however, the friction coefficient became consistent with Einstein-Sutherland equation only for mass ratios beyond \num{5e6}, while a deviation of about \SI{20}{\percent} were found for smaller masses. 

For the present system, the mass ratio between C\textsubscript{60} and a toluene molecule is about \num{8}. Thus, in line with the work of~\citet{ZoranThesis}, the integral of the \gls{ftacf} neither shows a plateau value, nor an exponential decay (cf. Figure~\zref{fig:results:facf}).
Instead the decay is governed by backscattering with a power law envelope consistent with the $t^{-5/2}$, corresponding to a proper hydrodynamic decay~\cite{Alder1970, Ould-Kaddour2003}. Here it is observed at relatively short timescales (inset in Figure~\zref{fig:results:facf}), due to the finite sampling times.

As a consequence of this behavior of the \gls{ftacf}, we are only left with the approximation of~\citet{Lagarkov1978}.
Accordingly, we obtain $\xi = \SI{3.37+-0.03e-12}{\kg\per\s}$ (corrected for finite size effects with equation~\zeqref{eq:friction-size}), which corresponds to a diffusion coefficient of $D = \SI{12.0 +- 0.1 e-10}{\m\squared\per\s}$, using the Einstein-Sutherland equation~\zeqref{eq:intro:einstein}.
Hereby, the error is the standard deviation between the simulations of different size after applying the size correction.

This obtained diffusion coefficient is about a factor of two larger than the diffusion coefficient obtained directly from the \gls{msd} or the \gls{vacf}. This result, furthermore, needs to be seen in the light of the obtained value for the friction which typically represents an overestimation of the friction coefficient~\cite{Kirkwood1946, Lagarkov1978, Espanol1993, Ould-Kaddour2003}.
As such, we clearly demonstrate the failure of equation~\zeqref{self:eq:friction:facf}, which is expected, given that the condition of a slow process underlying  equation~\zeqref{self:eq:friction:facf} is violated as demonstrated in section~\zref{sec:md}.

\subsubsection{Calculating the memory kernel}\zlabel{sec:si:memory-kernel}
\begin{figure}
    \includegraphics{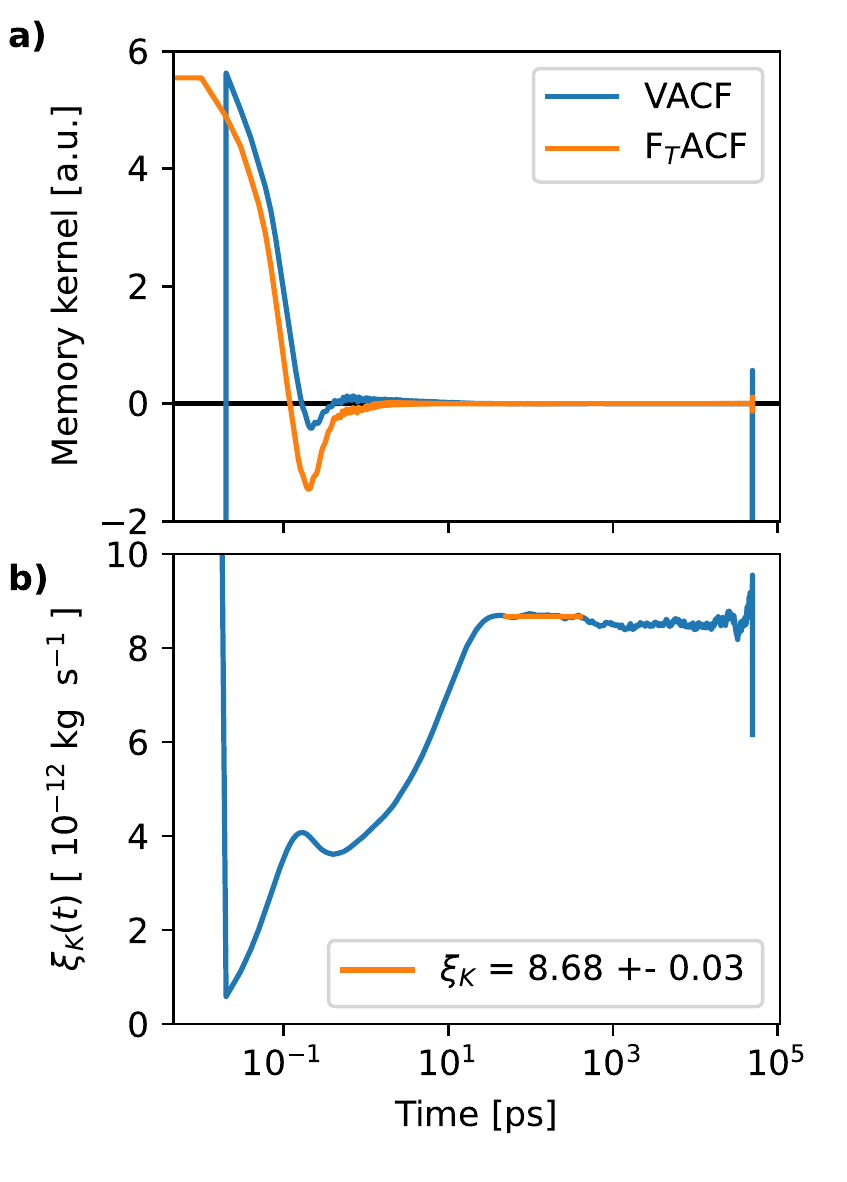}
	\caption{\textbf{Memory kernel of C\textsubscript{60} in toluene} for the system with 3824 toluene molecules.
	\textbf{a)} Memory kernel $K(s)$ calculated from the \gls{ftacf} directly and from the \gls{vacf} via the Fourier transform method (VACF) \cite{Kowalik2019, Gottwald2015}.
	\textbf{b)} Integral of the memory kernel derived from the \gls{vacf} with the FFT method. The integral saturates at the fitted value, which is derived in between \SIrange{10}{400}{\ps}.}
	\zlabel{fig:results:memory-kernel}
\end{figure}
The memory kernel $K(s)$ was by now calculated from the \gls{ftacf} as an approximation to the \gls{fsacf} (cf. SI section~\zref{sec:si:mori-zwanzig}).
Different methods to calculate the full memory kernel (without further approximations) have been proposed in the past years~\cite{Kowalik2019}, most of them using the position or velocity \glspl{acf} as base.
The Fourier method originally proposed by \citet{Gottwald2015} and summarized in \citet{Kowalik2019} is a straightforward and easy to apply method to calculate the kernel from the \gls{vacf}, which we apply using a fast Fourier transform method for calculating the (inverse) Fourier transforms.
We present the memory kernel derived from the \gls{ftacf} and from the \gls{vacf} via the Fourier method (FFT) in SI-Figure~\zref{fig:results:memory-kernel}a.
There are clear disagreements between the methods, demonstrating a failure of the approximation of the \gls{fsacf} with the \gls{ftacf}.

Integrating the memory kernel obtained from the \gls{vacf} with the Fourier method (cf. SI-Figure~\zref{fig:results:memory-kernel}b) yields $\xi_\textup{PBC} = \SI{8.68+-0.03 e-12}{\kg\per\s}$, which using the Einstein-Sutherland equation~\zeqref{eq:intro:einstein} corresponds to a diffusion coefficient of $D_\textup{PBC} = \SI{4.66+-0.02 e-10}{\m\squared\per\s}$, precisely the value obtained by integrating the \gls{vacf} directly.
Both values are for the same system and thus can be compared directly without correcting for system size effects.
The agreement is expected, as both methods are equivalent to obtain the diffusion coefficient, thus supporting the validity of the Einstein-Sutherland equation.

\subsubsection{D from NEMD simulations}\zlabel{sec:si:diffusion:pull}
The results of the previous section motivate us to calculate the friction coefficient using an alternative approach. Specifically, we use steered \gls{md} simulations to apply a constant force to the C\textsubscript{60} and measure its average translational velocity. The ratio of pull force and the resulting C\textsubscript{60} velocity should give the friction coefficient by equation~\zeqref{eq:intro:friction}.

This task is executed by applying a broad range of pull forces to the C\textsubscript{60} in a toluene box with side length \SI{13.8}{\nm} (see SI section~\zref{sec:si:methods} for details).
The regime of linear response is used as fit range for $\xi_\textup{PBS}$  (cf. Figure~\zref{fig:results:pull-friction}).
For each pull force, the relative velocity of the C\textsubscript{60} is calculated as the fullerene velocity, subtracted by the instantaneous fluid velocity, averaged over all fluid molecules.
For each simulation, the average is then calculated over the whole trajectory and the statistical uncertainty following the analysis of~\citet{AllenTildesly}.
Subsequently, the mean velocity is calculated as average over all replications and the statistical uncertainty with proper propagation of uncertainties.
At pull forces larger than \SI{50}{\kJ\per\mole\per\nm}, effects of non-linear response are evident, where non-equilibrium effects would have to be taken into account~\cite{Chu2019}.
Therefore, these simulations are left out of the analysis (green points in Figure~\zref{fig:results:pull-friction}).

Linear response is observed for forces smaller than \SI{50}{\kJ\per\mole\per\nm}.
In the range of \SIrange{10}{50}{\kJ\per\mole\per\nm}, we were able to properly converge the data, 
although still subject to fluctuations (orange points in Figure~\zref{fig:results:pull-friction}).
In this range, we find the friction coefficient to be \SI{7.5e-12}{\kg\per\s}, upon averaging over the entire range. The uncertainty of about \SI{2}{\percent} accounts for the inherent error of each data point and deviations within the interval. 

Within error bars, this result is identical to that at pull forces smaller than  \SI{10}{\kJ\per\mole\per\nm} (blue points in Figure~\zref{fig:results:pull-friction}). However, in this regime, the convergence of velocity is very slow. Consequently, the result is associated with large statistical uncertainties.
Notably, even these small forces are comparably large to those applied in sedimentation experiments. In fact, the maximum force applied in our \gls{auc} experiments and that of~\citet{Pearson2018} is about \SI{2e-6}{\kJ\per\mole\per\nm}. Obviously, these experiments are fully within the linear response regime, but with orders of magnitude larger observation times and particle numbers, such that proper sampling can be accomplished.
Therefore, experimental results should be comparable to simulations. 

We proceed to verify the validity of the Einstein equation. First, we account for the finite size effects using equation~\zeqref{eq:friction-size}, yielding $\xi = \SI{6.9+-0.15 e-12}{\kg\per\s}$.
This result is substantially different to that obtained from the \gls{ftacf}, contrary to previous reports. Namely, both~\citet{Espanol1993} and~\citet{Bocquet1994} performed simulations of a hard sphere fluid with a single particle fixed in space mimicking an infinitely heavy Brownian particle.
Differences between the friction coefficients evaluated from the fluid drag and the \gls{ftacf} were between \SI{4}{\percent} and \SI{8}{\percent} for different particle and system sizes.

Unlike in their system, where they emulate a particle with an infinite mass, in our system the molecular weight of C\textsubscript{60} is small.
As discussed previously, the separation of time scales is not achieved, hence the \gls{ftacf} does not yield a reliable result for $\xi$, and the two approaches to calculate the friction coefficient do not yield an even similar result.

Importantly, however, if $\xi$ obtained in steered \gls{md} simulations is used to calculate the diffusion coefficient, one obtains $D = \SI{5.9 +- 0.1 e-10}{\m\squared\per\s}$, with the statistical error of \SI{2}{\percent}. This is basically the same result as the diffusion coefficient obtained from the \gls{msd} or the \gls{vacf} (section \zref{sec:md} and SI section~\zref{sec:si:diffusion:vacf}).
The evaluation of the friction from a response to drag implies that the Einstein-Sutherland equation~\zeqref{eq:intro:einstein} holds for the case of C\textsubscript{60} in toluene.

\subsubsection{D from AUC}\zlabel{sec:si:diffusion:auc}
We furthermore perform \gls{auc} experiments to obtain diffusion and sedimentation coefficients $s$ and $D$, respectively.
We retrieved the mean sedimentation and diffusion coefficient of the main peak from the data analysis of the \gls{sv-auc} experiments using the $c(s)$ model in SEDFIT~\cite{Schuck2000} (a representative fit is shown in Figure~\zref{fig:AUC_fit}).
For a single peak $c(s)$-distribution, these values are independent of the \gls{psv}, as $s$ and $D_\textup{AUC}$ are directly fitted through the Lamm equation as long as solvent compressibility is not taken into account.
However, $s$ and $D_\textup{AUC}$ are both influenced by hydrodynamic and thermodynamic non-ideality effects.
Therefore, we conducted several \gls{sv-auc} experiments with varying C\textsubscript{60} mass concentrations and extracted the apparent coefficients at each finite mass concentration.
Extrapolation of the apparent values to the infinite dilution limit as presented in Figure~\zref{fig:results:AUC} is carried out to get a measure for the influence of concentration and to obtain $s$ and $D_\textup{AUC}$ at infinite dilution.

\begin{figure}[ht]
    \includegraphics[width=\columnwidth]{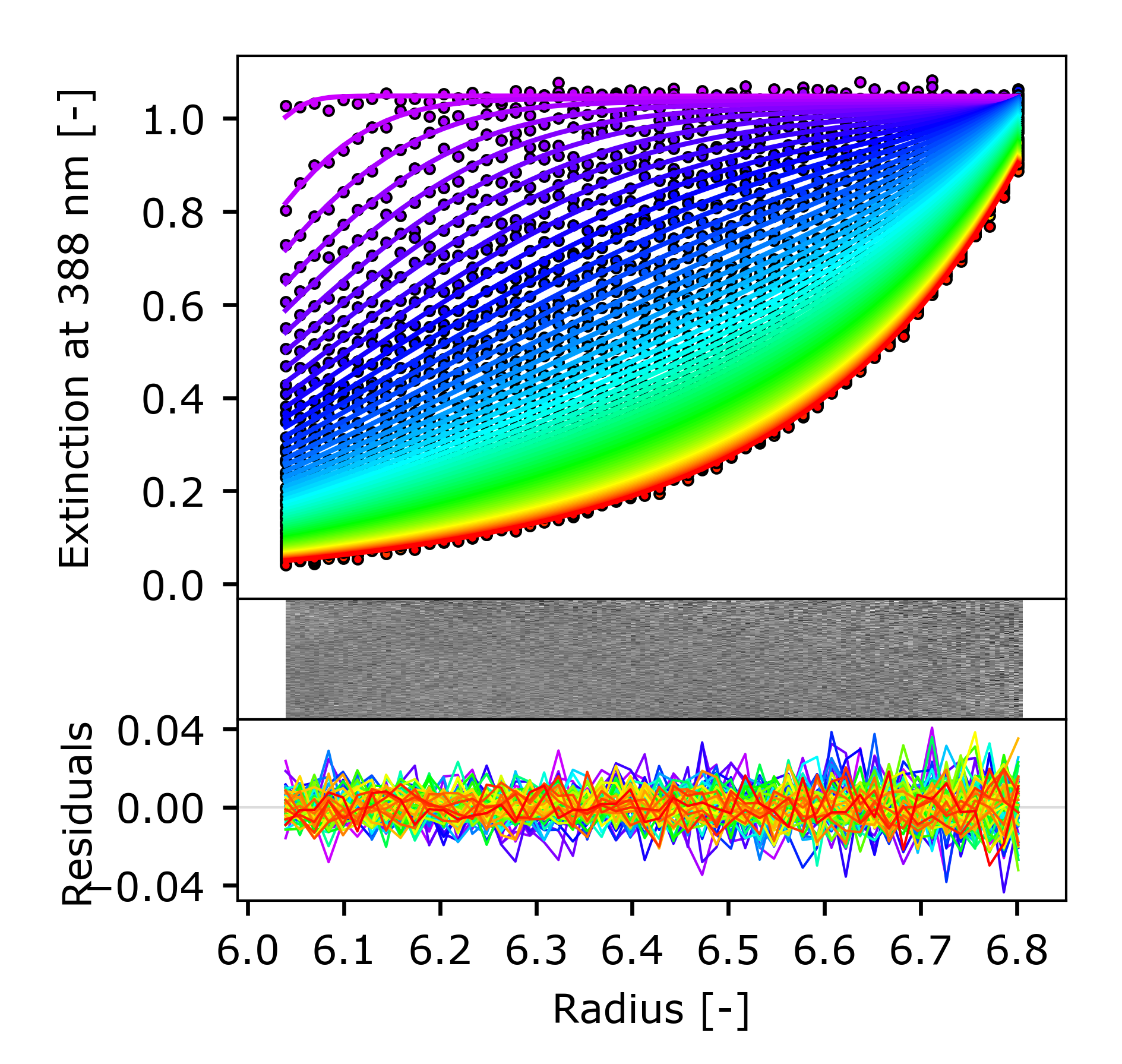}
	\caption{\textbf{Experimental data and fit for \gls{sv-auc} measurement of C\textsubscript{60} in toluene.} Experimental data (points) was recorded at $\SI{20}{\degreeCelsius}$, $\SI{60000}{rpm}$ and a loading concentration of \SI{0.08}{\g\per\L}. Best-fit sedimentation profiles (solid lines) were obtained by the c(s) inhomogeneous solvent model implemented in SEDFIT~\cite{Schuck2004}. The residuals between measured data and calculated sedimentation profiles are shown as map and lines below. Only every 6th scan and every 3rd data point is plotted. The plot was created using Gussi~\cite{Brautigam2015}.
	}
	\zlabel{fig:AUC_fit}
\end{figure}

The slope of the inverse apparent sedimentation coefficient versus mass concentration is a measure for hydrodynamic non-ideality~\cite{Uttinger2019}.
From our results in Figure~\zref{fig:results:AUC}a, it can be concluded that hydrodynamic non-ideality is not very pronounced and a linear fit of the inverse sedimentation coefficient describes the data well. Furthermore, the apparent diffusion coefficient is influenced by both, hydrodynamic and thermodynamic non-ideality, which is accounted for by the second virial coefficient~\cite{Harding1985, Uttinger2017}.
From Figure~\zref{fig:results:AUC}b, it can be concluded that thermodynamic non-ideality must be accounted for throughout data interpretation. The values at infinite dilution $s_0$ and $D_0$ are determined to be \SI{1.26 \pm 0.01}{sved} and $\SI{7.41+-0.04e-10}{\m\squared\per\s}$, respectively.
For an independent evaluation of the data, we conducted further experiments with a commercial Optima \gls{auc} and we also analyzed our \gls{sv-auc} data with SEDANAL, which yielded very similar results for $s$ and $D$ (details and results can be found in Appendix~\zref{sec:si:auc_evaluation}).

\subsection{Evaluations of the C\textsubscript{60} radius}
\zlabel{sec:si:radius:static}

Due to solvation effects, the apparent volume of a dispersed particle is not necessarily equal to its volume in the solid or the gas phase.
The change of particle size in solvents is captured by the partial molecular volume $V$, the latter calculated as the increase in volume of a system when suspending a single particle in fluid.
Typically, the changes are about \SIrange{10}{20}{\percent}.
For C\textsubscript{60} at infinite dilution $V$ was reported to be even smaller in a number of solvents compared to its crystal form.
However, for C\textsubscript{60} in toluene, $R_{\textup{V}}^{\textup{exp}} = \SI{0.524+-0.003}{\nm}$ has been obtained~\cite{Ruelle1996} (Table~\zref{tab:results:summary}), which is within the uncertainty of the measurements in vacuum.
Hereby, a perfect spherical shape of C\textsubscript{60} is assumed to deduce a radius from its volume.

To verify these results, we use our \gls{auc} experiments, where $V$ is  given as the product of the \gls{psv} and the molecular weight $M$, both of which are related to sedimentation and diffusion coefficients through the Svedberg equation~\zeqref{eq:Svedberg}.
Rearranging the latter to solve for the \gls{psv}, we obtain
\begin{equation}\zlabel{eq:psv-auc}
    \bar{\nu} = \frac{1}{\rho_\textup{S}} \left( 1 - \frac{s}{D} \frac{\kbt N_\textup{A}}{M} \right)
\end{equation}
Under the assumption of a uniform solvent density, multiplication of the \gls{psv} with the known mass of a C\textsubscript{60} molecule, we obtain the molecular volume from which we can then deduce its radius, again assuming a spherical shape.
Using the retrieved sedimentation and diffusion coefficients from  section~\zref{sec:si:diffusion:auc}, we obtain a radius $R_{\textup{V}}^{\textup{exp, AUC}} = \SI{0.520 +- 0.002}{\nm}$.
Within measurement uncertainties of less than \SI{1}{\percent} this equals the radius obtained by~\citet{Ruelle1996}.

Alternative techniques to determine the \gls{psv} and the corresponding radii experimentally are the Kratky method~\cite{Kratky1973} and the density contrast SE-AUC experiments. Both yield results in line with the above reported \gls{auc} data but with larger uncertainties:
\begin{align}
    R_V^\textup{exp, PSV} \textup{(Kratky, \SI{534}{\nm})} &= \SI{0.528+-0.007}{\nm} \\
    R_V^\textup{exp, PSV} \textup{(Kratky, \SI{570}{\nm})} &= \SI{0.495+-0.009}{\nm} \\
    R_V^\textup{exp, PSV} \textup{(DC SE-AUC)}             &= \SI{0.526+-0.014}{\nm}
\end{align}
The radius retrieved by SE-AUC is the closest to the radius calculated from the \gls{sv-auc} experiments.
The two values from the Kratky method originate from the optical determination of the particle concentration at two different wavelengths. The discrepancy between the two results is likely associated with the different, not fully consistent values for the mass extinction coefficient of C\textsubscript{60}, reported in the literature~\cite{Sension1991, Gunkin2006}.
Further details on the extinction coefficients and the determination of the \gls{psv} can be found in Appendix~\zref{sec:si:psv}).

\begin{figure}
\centering
\includegraphics{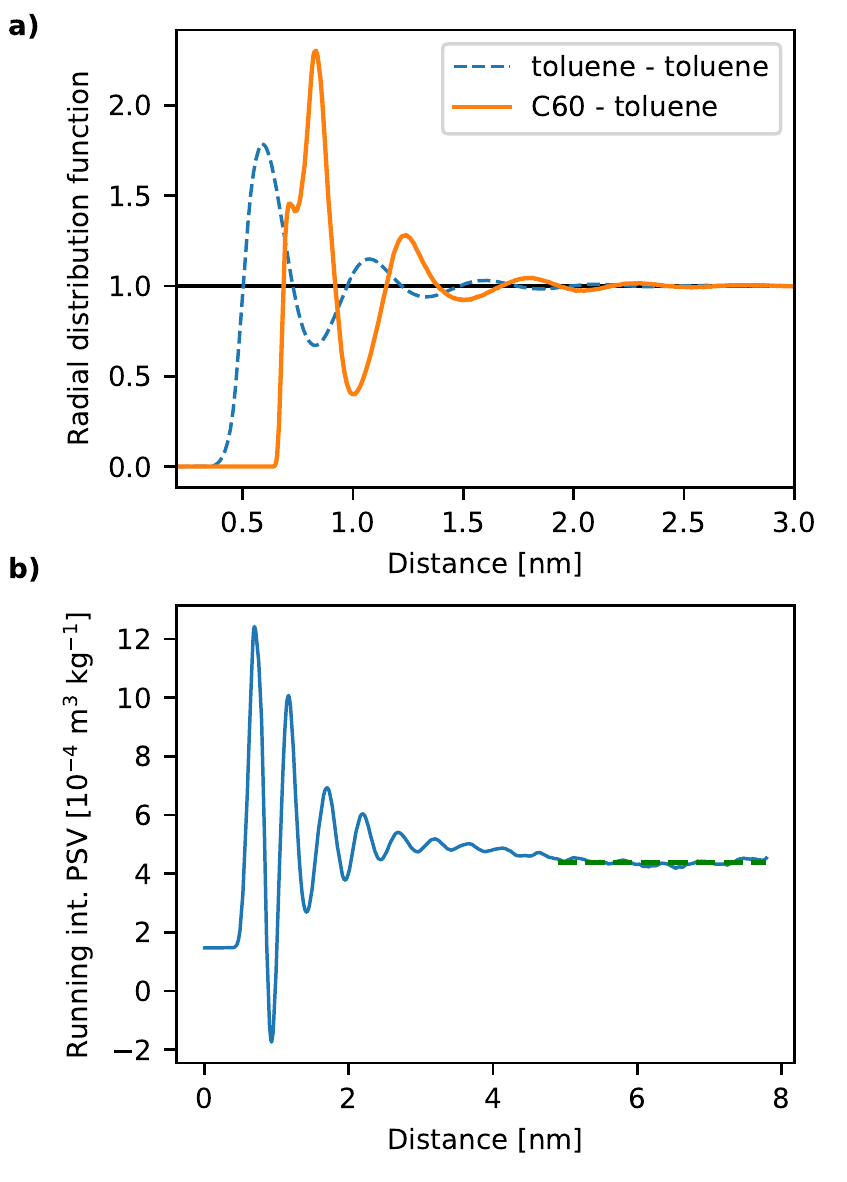}
\caption{\textbf{Structure of the toluene around C\textsubscript{60}.}
\textbf{a)}
\glspl{rdf} of toluene around C\textsubscript{60} ($g(r)$) and among toluene ($g_\textup{S}(r)$).
Displayed are the \glspl{rdf} between the centers of mass of the respective molecules.
\textbf{b)}
Running Kirkwood-Buff integral~\cite{Koga2013} to determine the \gls{psv} of C\textsubscript{60} in toluene.
With the MSER-5 algorithm, the point of convergence is determined and the \gls{psv} is then determined as the average of the data beyond that point (green dashed line).
}
\zlabel{fig:results:psv}
\end{figure}
In our \gls{md} simulations, we explicitly determine $g_\textup{s}(r)$ and $g(r)$ (cf. Figure~\zref{fig:results:psv}a) in the largest system, such that the convergence of the integral in equation~\zeqref{eq:radius-kb} is secured (see SI section~\zref{sec:si:methods} for details).
Hereby, we use the MSER-5 algorithm~\cite{MSER5} to find the most appropriate position for cutting the running integral.
$V$ is then determined as the average of the data beyond that point and the standard deviation in the regime serves as estimate for the statistical uncertainty (cf. Figure~\zref{fig:results:psv}b).
The result is $R_\textup{V}^\textup{{sim}} = \SI{0.500+-0.003}{\nm}$, which is only $4\%$ smaller than that found experimentally by~\citet{Ruelle1996}, and  $6\%$ smaller the radius directly form the force field. However, the fact that $R_{\textup{V}}^{\textup{sim}}$ is smaller than $R_{\textup{V}}^{\textup{exp}}$ suggests that the simulation force-field somewhat overestimates the interaction of C\textsubscript{60} and the toluene. Nonetheless, these deviations are small confirming the appropriateness of the molecular model of the fullerene and toluene for determining the static properties of the system, as expected.

\subsection{C\textsubscript{60} radius in gas and solid phase}\zlabel{sec:si:radius}
There are two distinct types of radii obtainable from static properties.
Firstly, the lengths of the two distinct C--C bonds can be measured.
From knowledge of the icosahedral structure~\cite{Johnson1990}, the radius of the full carbon structure can then be calculated.
In this case, the size of the C\textsubscript{60} shell needs to be taken into account explicitly.
Secondly, in the C\textsubscript{60} crystal the distance of nearest neighbors can be measured by various techniques.
For spheres in direct contact, the distance between nearest neighbors is equal to their diameter, thus, the radius can be retrieved.
All the resulting radii are summarized in Table~\zref{tab:lit:radii}.

\begin{table*}
\centering
\caption{Overview over radii obtained from different measurement techniques in \si{nm}.
The radii are:
$R_\textup{C}$: radius of the rigid carbon structure,
$R_\textup{S,min}$: $R_\textup{C}$ plus minimum intermolecular C$\cdot\cdot$C distance,
$R_\textup{S,g}$: $R_\textup{C}$ plus graphite interplanar distance,
$R_\textup{N}$: nearest neighbor distance in C\textsubscript{60} crystals.
$^\dag$ The distance measured is the planar centers of mass distance between neighboring C\textsubscript{60} in a C\textsubscript{60} monolayer. Due to the structure of the substrate, the individual fullerenes in the layer are at different heights, thus the measured distance is expected to underestimate the actual three dimensional distance (see text for details). Thus, this value is excluded from the average here.
}
\zlabel{tab:lit:radii}
\begin{ruledtabular}
\begin{tabular}{llll}
technique & {$R_\textup{C}$} & {$R_\textup{S,min}$} & {$R_\textup{S,g}$} \\
\hline
\glsentryshort{nmr}~\cite{Yannoni1991, Johnson1992} &
    0.355+-0.0035    & 0.5116+-0.0039 & 0.5225+-0.0035 \\
gas phase electron diffraction~\cite{Hedberg1991} &
    0.35565+-0.0005  & 0.5122+-0.0009 & 0.52315+-0.0005 \\
single crystal \glsentryshort{xrd}~\cite{Liu1991} &
    0.35325+-0.00015 & 0.5098+-0.0005 & 0.52075+-0.00015 \\
powder \glsentryshort{xrd}~\cite{Heiney1991} &
    0.352+-0.001     & 0.5086+-0.001  & 0.5195+-0.002 \\
powder \glsentryshort{xrd}~\cite{Stephens1991} &
    0.354+-0.002     & 0.5106+-0.002  & 0.5215+-0.002 \\
\hline
average & 0.3540+-0.0008 & 0.5106+-0.0009 & 0.5215+-0.0009 \\
\hline
\hline
technique && {$R_\textup{N}$} & {$R_\textup{V}$}\\
\hline
single crystal \glsentryshort{xrd}~\cite{Kraetschmer1990} && 0.501 & --- \\
\glsentryshort{hrtem}~\cite{Goel2004}                     && 0.505+-0.025 & --- \\
\glsentryshort{stm} (bulk crystal)~\cite{Wilson1990}      && 0.55 +-0.025 & --- \\
\glsentryshort{stm} (monolayer) on \ce{WO2}/W(110)~\cite{Murphy2014} && 0.475+-0.025$^\dag$ & --- \\
partial specific volume~\cite{Ruelle1996} && --- & 0.5243+-0.0026 \\
\hline
average && 0.519+-0.012 & 0.5243+-0.0026
\end{tabular}
\end{ruledtabular}
\end{table*}

\subsubsection{Bond lengths measurements}
\citet{Yannoni1991} and~\citet{Johnson1992} use \gls{nmr} spectroscopy to measure the bond lengths of the double bonds (edges between adjacent hexagons) $a_6 = \SI{0.140+-0.0015}{\nm}$ and of the single bonds (edges of a pentagon) $a_5 = \SI{0.145+-0.0015}{\nm}$.
A radius of $R_\textup{C} = \SI{0.355+-0.0035}{\nm}$ is calculated from this geometry.
\citet{Hedberg1991} used electron diffraction to retrieve bond lengths of $a_6 = \SI{0.1401+-0.0010}{\nm}$ and $a_5 = \SI{0.1458+-0.0006}{\nm}$, giving a radius of $R_\textup{C} = \SI{0.35565+-0.0005}{\nm}$.
\citet{Liu1991} did single-crystal \gls{xrd} experiments to obtain bond lengths of $a_6 = \SI{0.1355+-0.0009}{\nm}$ and $a_5 = \SI{0.1467+-0.0021}{\nm}$, resulting in an average radius of $R_\textup{C} = \SI{0.35325+-0.00015}{\nm}$.
\citet{Heiney1991} performed powder \gls{xrd} experiments at \SI{300}{\K}.
A fit to a theoretical curve gave a radius of $R_\textup{C} = \SI{0.352+-0.001}{\nm}$.
\citet{Stephens1991} performed powder \gls{xrd} of \ce{K3C\textsubscript{60}}, retrieving a diameter of the carbon cage of \SI{0.354}{\nm}.
The authors mention a distortion of the individual carbon atoms from the ideal positions of up to \SI{0.002}{\nm}, which they say to be small but significant.
All measurements are in excellent agreement with a relative deviation of only \SI{\approx 0.5}{\percent}, which is comparable to the measurement uncertainties.
The input structure used in the simulation is based on the \gls{nmr} measurement of~\citet{Yannoni1991, Johnson1992}.

As we aim for radii including the electron shell of C\textsubscript{60}, we need to account for it explicitly in this type of radius measurements.
It was suggested to take the interplanar distance of graphite layers, \SI{0.335}{\nm}, as estimate for the intermolecular C$\cdot\cdot$C distance, leading to an addition of \SI{0.1675}{\nm} to the radius~\cite{Dresselhaus1996}.
However, the authors also argue that the nominal $sp^2$ bonding differs from this in graphite as it occurs on a curved surface, leading to some admixture of $sp^3$ bonding.
\citet{Liu1991} also obtain from their \gls{xrd} measurements mostly values of about \SI{0.33}{\nm}, while some intermolecular distances found are as short as \SI{0.3131+-0.0007}{\nm}.
Using the graphite interplanar distance as estimate for the shell size and the radius of~\citet{Hedberg1991}, one obtains $R_\textup{S} = \SI{0.52315+-0.0005}{\nm}$.

\subsubsection{Nearest neighbors distance measurements}
The shorter than expected intermolecular distances hint to a slightly denser packing of C\textsubscript{60} than expected, which is further strengthened by other studies on crystalline C\textsubscript{60}.
\citet{Kraetschmer1990} conducted \gls{xrd} measurements of a pure C\textsubscript{60} crystal to obtain a nearest-neighbour distance of \SI{1.002}{\nm}.
Stating that the excess between this distance and the size of the carbon structure is the effective \gls{vdw} diameter of carbon in this molecule, they realized the larger than expected packing density resulting from this (cf. Table~\zref{tab:lit:radii} for details).
\citet{Goel2004} did \gls{hrtem} of a \SI{99.5}{\percent} pure C\textsubscript{60} crystal, directly placed on the TEM grid, to obtain center to center distances of \SI{1.01+-0.05}{\nm}, resulting in a radius of $R_\textup{N} = \SI{0.505+-0.025}{\nm}$.
\citet{Wilson1990} used \gls{stm} to measure the intermolecular distance of a C\textsubscript{60} crystal on a Au(111) surface.
A distance between the centers of \SI{1.10+-0.05}{\nm} was obtained, resulting in a radius of $R_\textup{N} = \SI{0.55+-0.025}{\nm}$.
Another study~\cite{Murphy2014} with an \glsentryshort{stm} of a C\textsubscript{60} monolayer on a \ce{WO2}/W(110) surface claims to obtain smaller intermolecular distances, but their result of \SI{0.95+-0.05}{\nm} is fully compatible with the other studies within measurement errors.
The author also mentions that the apparent height of the C\textsubscript{60} molecules differs significantly.
This is attributed to surface reconstructions and thus different types of binding sites for C\textsubscript{60}.
This out of plane arrangement decreases the measured planar distance between neighboring C\textsubscript{60} molecules, which is in full agreement with the other studies presented.


\bibliography{bibliography_no-eprint}